\documentclass[useAMS]{mn2e}
\twocolumn
\input{psfig}
\usepackage{times}
\usepackage{bm}
\usepackage{amsmath} 
\usepackage{url}
\usepackage{amssymb}
\usepackage{dblfnote}
\usepackage{aas_macros}
\usepackage{hyperref}

\def\Real{{\rm I\mathchoice{\kern-0.70mm}{\kern-0.70mm}{\kern-0.65mm}%
 {\kern-0.50mm}R}}

\def\bx#1{\leavevmode\thinspace\hbox{vrule\vtop{\vbox{\hrule\kern1pt
 \hbox{\vphantom{\tt/}\thinspace{\bf#1}\thinspace}}
 \kern1pt\hrule}\vrule}\thinspace}

\newcommand{\bea}{\begin{eqnarray}}
\newcommand{\eea}{\end{eqnarray}}
\newcommand{\be}{\begin{equation}}
\newcommand{\ee}{\end{equation}}

\def\be{\begin{equation}}
\def\ee{\end{equation}}

\def\ba{\begin{eqnarray}}
\def\ea{\end{eqnarray}}

{\catcode`\@=11}

\def\ltsima{$\; \buildrel < \over \sim \;$}
\def\lsim{\lower.5ex\hbox{\ltsima}}
\def\gtsima{$\; \buildrel > \over \sim \;$}
\def\gsim{\lower.5ex\hbox{\gtsima}}

\title[The effect of baryon physics on weak lensing
  tomography]{Quantifying the effect of baryon physics on weak
  lensing tomography}

\author[Semboloni et al.]{Elisabetta Semboloni$^{1}$\thanks{sembolon@strw.leidenuniv.nl},  Henk Hoekstra$^{1}$, Joop Schaye$^{1}$, Marcel~P. van Daalen$^{1,2}$,
\newauthor{Ian~G. McCarthy$^{3}$}\\
$^1$Leiden Observatory, Leiden University, P.O. Box 9513, 2300 RA , Leiden, The Netherlands \\
$^2$Max Planck Institute for Astrophysics,  Karl-Schwarzschild Stra\ss{}e 1, 85741 Garching, Germany \\
$^3$Kavli Institute for Cosmology, University of Cambridge, Madingley Road, Cambridge CB3 0HA, United Kingdom\\
}
\begin{document}
\maketitle
\begin{abstract}
We use matter power spectra from cosmological hydrodynamic simulations  to quantify the
effect of baryon physics on the weak gravitational lensing shear
signal. The simulations  consider a number of processes, such as radiative cooling,
star formation, supernovae and feedback from active galactic nuclei
(AGN). Van Daalen et al. (2011) used the same simulations to show  that baryon physics, in particular the strong feedback that is required to solve the overcooling problem,  modifies the matter power
spectrum on scales relevant for cosmological weak lensing studies.  As a result, the use of power spectra from dark
matter simulations can lead to significant biases in the inferred
cosmological parameters.  We show that the typical biases are much larger than the
precision with which future missions  aim to constrain the dark energy
equation of state, $w_0$. For instance, the simulation with AGN
feedback, which reproduces X-ray and optical properties of
groups of galaxies, gives rise to a $\sim 40\%$ bias in $w_0$. 
We also explore the effect of baryon physics on  constraints on $\Omega_{\rm m}$, $\sigma_8$, the running of the
spectral index, the mass of the neutrinos and models of warm dark matter. We demonstrate that the
modification of the power spectrum is dominated by groups and clusters
of galaxies, the effect of which can be modelled.  We consider an
approach based on the popular halo model and show that simple
modifications can capture the main features of baryonic feedback.
Despite its simplicity, we find that our model, when calibrated on the 
simulations, is able to reduce the
bias in $w_0$ to a level comparable to the size of the statistical
uncertainties for a Euclid-like mission. While observations of the gas and
stellar fractions as a function of halo mass can be used to calibrate
the model, hydrodynamic simulations will likely still be needed to
extend the observed scaling relations down to halo masses of $10 ^{12}\,h^{-1}\, M_\odot$.
\end{abstract}


\section{Introduction}

The discovery that the expansion of the Universe is accelerating is
arguably one of the most significant discoveries of modern cosmology.
We lack, however, a theoretical framework to explain the
observations. For instance, current measurements can be explained by a
cosmological constant or dynamic mechanisms (e.g., quintessence).  The
field driving the acceleration has been dubbed ``dark energy'', which
can be parametrised by an equation-of-state $w_0$ (and additional
evolving parameters). Modifications of the laws of gravity on
cosmological scales have been considered as well (for a recent review, see for example Uzan 2007).

Observations of type Ia supernovae  provided the first evidence for the accelerated expansion  \cite{Peetal99,Rietal07,Ametal10}.
Further support has come from observations of the cosmic microwave
background (e.g., Komatsu et al.\ 2011),  baryon acoustic oscillations
(e.g., Eisenstein et al.\ 2005; Percival et al.\ 2010) and gas fractions of massive clusters  (e.g., Allen et al.\ 2008). These methods allow us
to constrain the properties of dark energy solely by studying the expansion
history of the Universe and  do not
distinguish between modified theories of gravity that  do not alter the
expansion history. 

Studies of the growth of structure are sensitive to both dark energy
and modifications of gravity, and are therefore of great interest.  The
interpretation of the observations is complicated by the fact that
most of the matter is in the form of dark matter. Fortunately,
inhomogeneities of the matter distribution cause the light of
background sources to be differentially deflected. This leads to
coherent alignments in the shapes of these galaxies, which can be
related directly to the power spectrum of matter density fluctuations.
This phenomenon is referred to as ``weak gravitational lensing'' and
when applied to the study of large-scale structure, it is also   known as
``cosmic shear''.

The cosmic shear signal has been measured using large ground-based
surveys (e.g., Hoekstra et al.\ 2006; Fu et al.\ 2008) and from space
(e.g., Massey et al.\ 2007; Schrabback et al.\ 2010).  As discussed by
the Dark Energy Task Force (DETF; Albrecht et al.\ 2006), it  is one of
the most promising probes of dark energy. Space-based missions are
currently planned with the aim to constrain the dark energy
equation of state with a relative precision of $\sim 1\%$. The
interpretation of the signal will, however, require exquisite knowledge of
the power spectrum, well into the non-linear regime.

In order to exploit the statistical power of these future data sets,
we need to be able to model the power spectrum with percent
accuracy. To date, the interpretation of cosmic shear measurements
rely on results based on N-body simulations that contain only
collisionless (dark matter) particles. Under the assumption that the
evolution of the large-scale structure is mainly described by the
evolution of the dark matter distribution, this should lead to fairly
accurate results.  Recently, Heitmann at al. (2010) developed an
`emulator' which is able to provide dark matter only models with
 $1\%$ accuracy on scales larger than a megaparsec.

However, a complete description of the matter fluctuations should also include baryon physics.  Bernstein (2009) suggests to account for the effect of baryons on two-point weak lensing measurements  by adding an unknown component to the power spectrum which is expressed using Legendre polynomials. The nuisance of the baryons  is then included in the figure of merit (Albrecht et al. 2006) by marginalising over this component. Similarly, Kitching \& Taylor (2010) propose a path-marginalisation  technique and claim that this procedure  reduces the figure of merit  by  $10\%$. In contrast, Zentner, Rudd \& Hu (2008) suggest that constraining baryonic feedback and cosmology at the same time would degrade the figure of merit by about $30\%-40\%$. From those studies it emerges that self-calibrating techniques are very promising although the strength of the result relies on prior assumptions; because of that  it seems clear  that the better we understand how various mechanisms  affect the density distribution of the baryons,  the better we can choose priors in the self-calibration procedures. For this reason it is important to study the effect of baryonic feedback  directly using hydrodynamic simulations and this is the approach we adopt in this paper. 

 Hydrodynamic simulations are much more
expensive than collisionless N-body simulations, and it has only
recently become possible to simulate cosmological volumes with
reasonable resolution. These simulations aim mainly to reproduce the
properties of galaxies and the intergalactic medium, and to do that
they need to account for processes such as radiative cooling, star
formation, stellar mass loss, chemical enrichment and outflows
driven by supernovae and AGN. To complicate the situation further, it is not yet clear which processes one needs to include in the
simulations to mimic the observed Universe and the simulations still
lack the resolution to model all relevant processes from first
principles.
The OverWhelmingly Large Simulations (OWLS) project \cite{Schayeetal10}
represents an effort in that direction. The main goal of the OWLS
project is to study how the properties of large-scale structures,
galaxies and the intergalactic medium change in different scenarios, by producing a
large number of simulations and by  varying parameters independently.
Comparisons with observations allow one to establish which
mechanisms are required to better describe our Universe.  One of the main
results of the OWLS project is that the simple insertion of baryons,
star formation and feedback from supernovae, cannot reproduce the characteristics of
groups of galaxies. Accounting for AGN feedback, however, results in 
good agreement with both X-ray and optical observations (McCarthy et al.\ 2010).

Importantly for cosmic shear studies, van Daalen et al.\ (2011) used the OWLS runs to show that the resulting power spectrum of matter fluctuations
depends strongly on the adopted baryon physics out to surprisingly large scales. In
particular, their model for AGN feedback leads to significant changes
compared to scenarios with only star formation and relatively inefficient feedback from supernovae. In this latter
scenario, cooling baryons change the profiles of the haloes, generating
more compact cores and increasing power on small scales (e.g.,\ Jing et al.\ 2006; Rudd et al.\ 2008, Guillet et al.\ 2010; Casarini et al.\ 2011). To prevent the formation of overly luminous
galaxies and to reproduce the X-ray properties of groups, extremely efficient feedback is required. AGN feedback can prevent the overcooling of the baryonic component by ejecting large quantities of gas at high redshift, when the supermassive black holes were growing rapidly (McCarthy et al.\ 2010; 2011). Consequently,
AGN feedback can modify the power spectrum in a dramatic way, decreasing the scale below which cooling increases the power, and significantly reducing the power up to very large scales.  Levine \& Gnedin (2006)  used a simple  toy model to manually include AGN outflows in collisionless simulations and to show that AGN can potentially drastically change the power spectrum, although they were unable to predict the sign and the magnitude of the effect.
More recently, van Daalen et al.\ (2011) predicted, using a fully hydrodynamic simulation that reproduces optical and X-ray observations of groups of galaxies,  a $\sim 10$\% decrease in total power relative
to a dark matter only simulation on scales of $0.1 - 10~  h^{-1}~{\rm Mpc}$ at $z=0$.

Ignoring the effects of baryon physics may therefore lead to biases in
the constraints of cosmological parameters inferred from cosmic shear
studies. In this paper we use the power spectra tabulated by van Daalen et al.\ (2011) to evaluate the effects of a number of feedback
scenarios. In addition, we provide a simple, yet rather effective approach to
reduce the bias of cosmological parameters.  We show that this kind of approach will be crucial for upcoming surveys because  the bias will be much larger than the statistical uncertainties, if left uncorrected.

This paper is organised as
follows. In section \ref{sec:OWLS} we describe the set of simulations
we use for this work. In section \ref{sec:2pt} we examine how cosmic
shear statistics are affected. In section \ref{sec:results} we
quantify the biases introduced when ignoring the effects of baryon
physics. In section \ref{sec:toy_model} we show that it is possible
to recover realistic power spectra and to remove (most of) the biases by making use of simple models that are calibrated to match the star and gas fractions of the simulated groups.
Cosmic shear cannot only constrain the properties of the dark energy, but can also be used to study  the  running of the spectral index, warm dark matter and massive neutrinos. In section \ref{sec:parameters}  we evaluate the consequences of baryonic feedback on the ability of cosmic shear  to constrain those effects.
Finally, we conclude in section \ref{sec:conclusions}.

\section{Simulations}\label{sec:OWLS}

To examine the physics that drives the formation of galaxies and the evolution of the intergalactic medium, 
the OverWhelmingly Large Simulations (OWLS) project includes over 50
large, cosmological, hydrodynamic simulations run with a modified version of the SPH code Gadget (last described in Springel 2005). A range of physical
processes was considered, as well as a range of model parameters. In
this paper we use a subset of OWLS for which we will give a brief
description, inviting the reader to find more details in the papers
where the simulations are presented. Note that all simulations have
been performed with the same initial conditions\footnote{The cosmology
  used to realise the simulations is the best-fit to the WMAP3 data
  \cite{Spetal07}: $\{\Omega_{\rm m},\Omega_{\rm b},
  \Omega_\Lambda,\sigma_8,n_s,h\}=\{0.238,0.0418,0.762,0.74,0.951,0.73\}$.}.
The simulations considered here are (following the naming convention
of Schaye et al.\ 2010):

\begin{itemize}
\item DMONLY: a dark matter only simulation, of the kind
   commonly used to compute the non-linear power spectrum  which is needed
  in weak lensing studies. It is therefore the reference to which we
  compare the other simulations. 
\item REF: although it is not the reference simulation for the study
  presented here, this simulation includes most of the mechanisms which
  are thought to be important for the star formation history (see Schaye et
  al. 2010 for a detailed discussion) but not AGN feedback. The implementation of radiative cooling, star formation, supernovae driven winds, and stellar evolution and mass loss have been described in Wiersma, Schaye \& Smith
  (2009), Schaye \& Dalla Vecchia (2008), Dalla Vecchia \& Schaye (2008), and Wiersma et al.\ (2009), respectively.  This simulation represents a standard
  scenario assumed in cosmological hydrodynamic simulations. 
\item DBLIMFV1618: this simulation has been produced using the same mechanisms as REF.  The only difference between the two simulations is that  in this simulation the stellar initial mass function (IMF) 
  was modified to produce more massive stars when the pressure of the gas
  is high, i.e.\ in starburst galaxies and close to galactic
  centres.  This is obtained by switching from the Chabrier (2003) IMF assumed in the REF model to a Baugh et al.\ (2005) IMF in those regions.
There are both observational and and theoretical arguments to support a top-heavy IMF in those extreme conditions (e.g. Padoan et al.\ 1997; Baugh et al.\ 2005; Klessen et al.\ 2007; Maness et al.\ 2007; Dabringhausen et al.\ 2009; Bartko et al.\ 2010; Weidner et al.\ 2010). The IMF change  causes the number of supernovae and the effect of
  stellar winds to increase resulting in a suppression of the SFR at smaller redshifts. However, this mechanism alone is not able to
  reproduce the observed SFR (see Schaye et al.\ 2010).
\item AGN:  a hydrodynamic simulation which differs from REF only by the inclusion of AGN. 
 The AGN feedback
  has been modelled following Booth \& Schaye (2009). In this approach
  AGN transfer energy to the neighbouring gas, heating it up and
  driving supersonic outflows which are able to displace a large
  quantity of baryons far from the AGN itself. Among the three
  simulations considered here, it is arguably the most realistic, as
  it is able to reproduce the gas density, temperature, entropy, and
  metallicity profiles inferred from X-ray observations, as well as
  the stellar masses, star formation rates, and stellar age
  distributions inferred from optical observations of low-redshift
  groups of galaxies \cite{MCetal10}.
\end{itemize}
 
To forecast the cosmic shear signal for the four different scenarios,
we make use of the results of van Daalen et al.\ (2011), who tabulated
the power spectra of matter fluctuations $P(k,z)$ in redshift slices
over the redshift range $0\leq z\leq 6$ for a number of OWLS runs. A
detailed discussion of the procedure to compute the power spectrum and
its accuracy can be found in van Daalen et al.\ (2011). Their
convergence tests and noise estimations suggest that the power
spectra estimated from OWLS are reliable up to at least $k \approx
10\, h\,{\rm Mpc}^{-1}$ over the range of redshifts we are interested
in (i.e. $z \lesssim 1$, as the lensing signal is most sensitive to
structures that are halfway between the observer and the source).  At
small $k$, the estimate of the power spectrum is affected by the
finite size of the simulation box ($100\, h^{-1}\, {\rm Mpc}$ on a
side).  This is not a concern, because on these scales baryonic
effects are very small and density fluctuations are in the linear
regime, so that we can compute the power spectrum from theory instead.

\begin{figure}
\psfig{figure=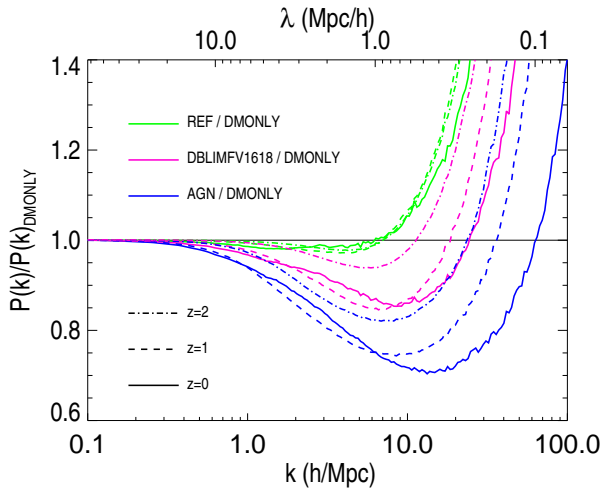,width=.45\textwidth}
\caption{\label{fig:Pkappa} Ratio between the power spectrum of matter
  fluctuations measured from the simulations with baryons and the one
  measured from the DMONLY simulation. The ratio for the REF
  simulation is shown in green, the one for the AGN simulation is
  shown in blue, and the one for the DBLIMFV1618 model is shown in
  pink. Since the simulations have been carried out using the same
  initial conditions, deviations of the ratio from unity are due to
  the differences in baryon physics.}
\end{figure}

Figure \ref{fig:Pkappa} shows the power spectrum measured for each
simulation in three redshift bins normalised by the power spectrum of
the dark matter simulation (DMONLY) at the same redshift.  In the REF
scenario (green), the presence of the baryons slightly suppresses the
power spectrum at intermediate scales, due to the pressure of the gas. At smaller
scales where baryons cool, the power spectrum is enhanced as the
baryons fall into the potential wells. For this model, only the small
scales are affected in an almost redshift independent way.  The effect
of baryon physics is more pronounced for the DBLIMFV1618 model, and
depends on redshift. The AGN model leads to the largest difference
compared to the DMONLY simulation. The amplitude of the power spectrum
is strongly reduced on scales of $\sim 1-10\,h^{-1}$ Mpc and the effect
increases as the redshift decreases; this is in agreement with  the results by McCarthy et al.\ (2011) who showed that because AGN remove low-entropy gas  at  early stages ($2 \lesssim z \lesssim 4$), the high-entropy gas left in the haloes does not cool down and form stars and the suppression  of power becomes more and more accentuated at small scales.

 The latter two scenarios are qualitatively similar, although the mechanisms are different: in the DBLIMFV1618 simulation baryons are removed due to
the enhanced supernova feedback, whereas in the AGN scenario they are
removed mostly by AGN feedback, at least for the more massive and thus strongly clustered haloes. Thus, the fraction of baryons which is removed is different, as is  the rate at which they are removed. At very small scales  cooling still enhances
structure formation, but the physical scale below which this occurs is
smaller than in the REF simulation.

Although  we will focus on a set of simulations which have been produced using the WMAP3 cosmology, van Daalen et al.\ (2011) compared a dark matter only simulation and an AGN simulation realised with the best-fit WMAP7 cosmology (Komatsu et al.\ 2011) and found  that  the difference is the same as between the AGN and DMONLY simulations used here.  This implies that the relative effect of baryonic feedback does not change significantly with the cosmology, and thus that our conclusions are not restricted to a specific set of cosmological parameters.

\begin{figure}
\psfig{figure=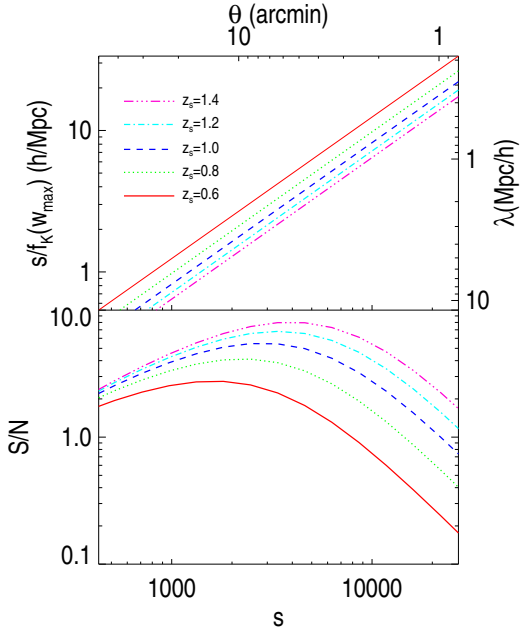,width=.45\textwidth}
\caption{\label{fig:distances} Top panel: relation between the wave number $k=s/f_K(w_{\rm max})$ and angular wave number   $s$ for various source redshifts. This relation shows, for a
  given source redshift, which  $k$ contributes most to the convergence power spectrum  at a given $s$. For both $s$ and $k$ we show the corresponding real-space conjugate variables.
Bottom panel:
  typical $P_\kappa(s)$ signal-to-noise ratio as a
  function of the  wave number   $s$. We show results for the same set of source redshifts as were shown in
  the upper panel.}
\end{figure} 
 
\section{Effect of baryons on two-point shear statistics}\label{sec:2pt}

In this section we briefly introduce the basics of weak gravitational
lensing by large-scale structure and how baryon physics affects the
interpretation of the measurements. For a more extensive review of 
 cosmic shear, see for example Hoekstra \& Jain (2008) and
Munshi et al.\ (2008).  A thorough discussion of the theory of weak
lensing and its applications is given in Bartelmann \& Schneider (2001).

Massive structures along the line of sight deflect photons emitted by
distant galaxies. Provided the source is small, the effect is a
remapping of the source's surface brightness distribution: the source
is both (de)magnified and sheared. In the weak lensing regime, the
convergence $\kappa$ gives the magnification (increase in size) of an
image and the shear $\bm{\gamma}$ gives the ellipticity induced on an
initially circular image. Under the assumption that galaxies are
randomly oriented in the absence of lensing, the strength of the tidal
gravitational field can be inferred from the measured ellipticities of
an ensemble of sources.  The resulting complex shear
$\bm{\gamma}\equiv\gamma_1+i\gamma_2$ is a spin-2 pseudo vector, which
can also be written as $\bm{\gamma}= \gamma\ {\rm exp}(2i\alpha)$,
where $\alpha$ is the position angle of the shear.

If redshift information is available for the sources, one can compute the two-point shear correlation functions for galaxies in
the $i^{\mathrm th}$ and $j^{\mathrm th}$ redshift bins, which are defined as

\begin{eqnarray}
\xi_{+}^{i,j}(\theta)=\langle\gamma_{i,\mathrm t}(\bm{\theta}_1)\gamma_{j,\mathrm t}(\bm{\theta}_2)+\gamma_{i,\times}(\bm{\theta}_1)\gamma_{j,\times}(\bm{\theta}_2)\rangle\\
\xi_{-}^{i,j}(\theta)=\langle\gamma_{i,\mathrm t}(\bm{\theta}_1)\gamma_{j,\mathrm t}(\bm{\theta}_2)-\gamma_{i,\times}(\bm{\theta}_1)\gamma_{j,\times}(\bm{\theta}_2)\rangle
\label{eqn:shearcorrelation}
\end{eqnarray}

\noindent where $\theta = |\bm{\theta}_1 - \bm{\theta}_2|$ and we  have defined the tangential and cross components of the shear as: $\gamma_{\mathrm t}=-{\mathcal Re}(\gamma\exp(-2i\phi))$ and $\gamma_\times=-{\mathcal Im}(\gamma\exp(-2i\phi))$, with $\phi$ the polar angle of the separation vector $\bm{\theta}$. Note that the ensemble average depends only on the  angular distance between the galaxies. The measurement of the redshift dependence of the
cosmic shear signal greatly improves the constraints on cosmological
parameters from weak lensing and is a key goal of current  and future surveys.
It is often referred to as weak lensing tomography (e.g.\ Hu 1999; Hu 2002), because it allows us to study the matter distribution in ``slices''. The correlation functions can be
measured from a catalogue of galaxy shapes  and they are related to 
 the convergence cross-power spectrum:

\begin{equation}
\xi_{+/-}^{ij}(\theta)=\frac{1}{2\pi}\int_0^\infty {\rm  d} s^\prime s^\prime {\rm J}_{0/4}(s^\prime \theta)
P_\kappa^{ij}(s^\prime),
\end{equation}

\noindent where $J_0$, $J_4$ are  the $0^{th}$ and the $4^{th}$ order Bessel functions of the first
8kind.  The convergence
power spectrum $P_\kappa^{ij}(s)$ is related to the power spectrum of
matter fluctuations $P(k,w)$ through \cite{Ka98,Scetal98}:

\be\label{eq:pkappa}
P^{ij}_\kappa(s)=\frac{9H_0^4}{4c^4} \Omega_{\rm m}^2 \int^{w_{\rm H}}_{0} {\rm d}w
\frac{g_i(w)g_j(w)}{a^2(w)} P\Big(\frac{s}{f_\mathrm{K}(w)},w\Big), 
\ee

\noindent with

\be\label{eq:distr}
g_i(w)=\int_w^{w_{\rm H}} {\rm d} w^\prime p_i( w^\prime)
\frac{f_\mathrm{K}(w^\prime-w)}{f_\mathrm{K}(w^\prime)}\;,  
\ee

\noindent where  $f_\mathrm{K}(w)$ is the comoving angular distance, $w$ is the radial comoving coordinate, $w_{\rm H}$ is the radial comoving coordinate of the horizon; $H_0$,
$\Omega_{\rm m}$ and $a(w)$ are the Hubble constant, the matter
density parameter and the scale factor, respectively. The projected
power spectrum depends on $p_i(w)$, the radial distribution of the
sources in the $i^{\mathrm th}$ redshift bin.  Since $s$ is the Fourier-conjugate of the angle $\theta$, we can relate an angular scale to it through  $s= 2\pi/\theta$.
  
Various two-point shear statistics  that  have been employed in the literature  simply correspond to different filters of $P_\kappa(s)$ in
Eqn.~\ref{eq:pkappa}. Therefore, the detailed effect of baryonic feedback
will depend somewhat on the statistic that is used, but this does not significantly affect our conclusions which are derived   using  $P_\kappa(s)$. 
Note that in the case of a full sky survey one could determine the $C(l)$ from a decomposition in spherical harmonics. For small angular separations $s \approx l$ and $l(l+1)C(l) \approx s^2P_\kappa(s)$.

\subsection{Relevant scales}\label{sub:scales}
Cosmic shear results are typically presented in terms of angular
scales, rather than the physical scales  shown in
Fig. \ref{fig:Pkappa}. We therefore start by examining which angular
scales are affected by baryon physics.  To do so, we consider sources
at a single redshift. In this case one can show that the value of  $P_\kappa(s)$  depends mostly on the density fluctuations with comoving wave numbers $ \approx s/f_K(w_{\rm max}) $ with $f_K(w_{\rm max})$ maximising the ratio
${\frac{f_\mathrm{K}(w_s-w)f_\mathrm{K}(w)}{f_\mathrm{K}(w_s)}}$. In the top panel of Figure \ref{fig:distances}, we show, for various
source redshifts $z_s$, the relation between the angular wave number  $s$
and the  wave number $s/f_K(w_{\rm max})$ using the adopted WMAP3
cosmology. It shows, for example, that measuring the power spectrum $P_\kappa(s)$ at $s \sim 1\times 10^4$
 of galaxies with redshifts $\sim 0.8$,  probes density fluctuations at scales  $k \sim 10 h\,{\rm
  Mpc}^{-1}$,  where baryon physics is important.

However, one might wonder if the signal at arcminute scales is
statistically important. To examine this, the bottom panel of Figure
\ref{fig:distances} shows a typical  signal-to-noise ratio of $P_\kappa(s)$.  The signal has been computed assuming a WMAP3 cosmology. The noise accounts for sampling and statistical noise, assuming a WMAP3 cosmology and a survey area  $A=20000~ {\rm deg}^2$, a number density of
galaxies of $n=30~ {\rm gal/arcmin^2}$ all  placed at the same redshift $z_s$ and with intrinsic ellipticity dispersion $\sigma_e=0.33$ (see  section \ref{sec:results} for more details on the noise computation).  As one can see, the signal-to-noise ratio peaks at scales between $2$ and $10$ arcmin,
where  baryon physics is important.

\begin{figure}
\psfig{figure=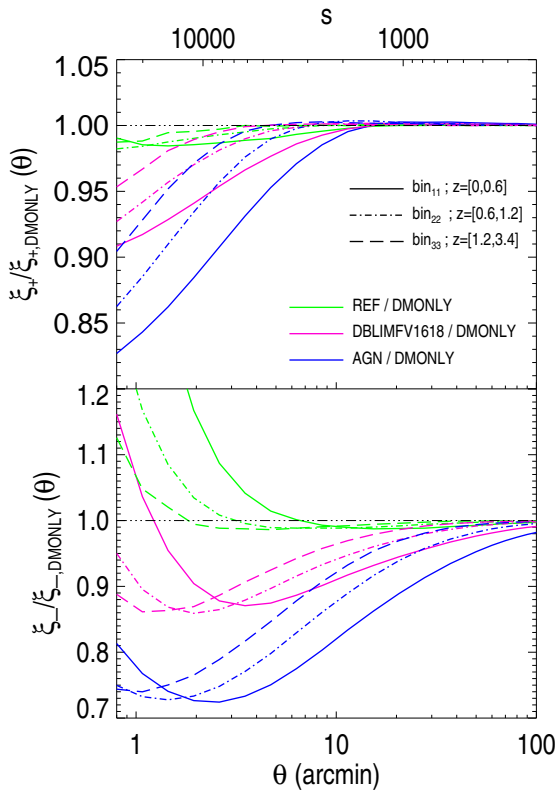,width=.45\textwidth}
\caption{\label{fig:gamma} Top panel: ratio  of the
  correlation function $\xi_+(\theta)$ for REF/DMONLY
  (green), DBLIMFV1618/DMONLY (pink) and AGN/DMONLY (blue) .  The notation
  ${\rm bin}_{ij}$ indicates the correlation of sources from 
  redshift bin $i$ with sources from redshift bin $j$. Here, we show
  only results from the bins with $i=j$. Bottom panel: same as the upper panel but for the correlation function $\xi_-(\theta)$.}
\end{figure}  

Having established that cosmic shear studies are sensitive to the
scales where baryon physics modifies the power spectrum, we now want
to quantify how various scenarios change the two-point shear
statistics. For that we adopt a source redshift distribution that is
representative of the CFHTLS-Wide \cite{Beetal07} and a fair
approximation for Euclid \cite{La09}. We adopt the following parametrisation:

\begin{equation}\label{eq:ps}
p(z)=\frac{\alpha}{(z+z_0)^\beta}\;,
\end{equation}

\noindent with $\alpha=0.836$, $\beta=3.425$, and $z_0=1.171$. We
divide the source galaxies in three tomographic bins with limits [0, 0.6, 1.2, 3.4],
which yields six cross-power spectra.

The top panel of Figure~\ref{fig:gamma} shows the  value of
$\xi_+(\theta)$    measured for the various feedback scenarios,
normalised by the results for DMONLY. The effect of
baryons is small and limited to very small scales for the REF
scenario. However, for DBLIMFV1618, and in particular for the AGN
model, the difference with the DMONLY result is large and increases
when the redshift of the sources decreases. The redshift dependence is the result of two
effects. The first is a geometric one: when the redshift of the
sources decreases, the physical scales probed by the lensing signal
 become smaller (see Figure \ref{fig:distances}). The second
reason is the suppression of the amplitude of the power spectrum due
to feedback, which becomes larger at late times (see Figure~\ref{fig:Pkappa}). The bottom panel of Figure~\ref{fig:gamma} shows the  value of
$\xi_-(\theta)$   measured for the various feedback scenarios, normalised by the results for DMONLY.  Notice that the bias for $\xi_-$ is more pronounced out to larger scales. This is because $\xi_-$ is much more sensitive to small-scale structures (i.e. to the shape of the power spectrum $P_\kappa(s)$ for large $s$). 

\begin{figure*}
\psfig{figure=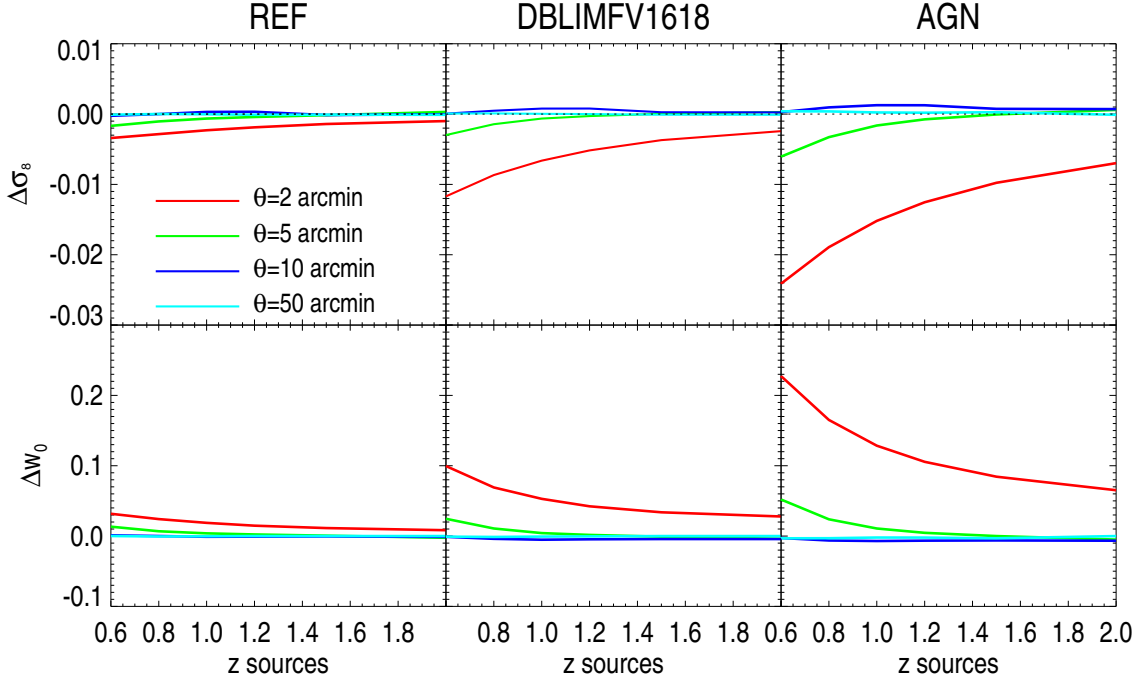,width=.98\textwidth}
\caption{\label{fig:delta_w} Top (bottom) panels show the deviation of the inferred 
  $\sigma_8$ ($w_0$) from the true reference value $\sigma_{8, \rm ref}=0.74$
  ($w_{0,\rm ref}=-1$) as a function of source redshift, when the amplitude of
  the ellipticity correlation function $\xi_+(\theta)$ is used to estimate the
  cosmological parameter of interest (while the other parameters
  are kept at their reference values) and when we use halofit models (see text for details). The deviation depends on the
  angular scales that is used and is smaller for larger scales. The
  left panels show the results for the REF scenario, the middle panels for the  DBLIMFV1618 and the right panels for the AGN  scenario, which results in the largest biases.}
\end{figure*}  

\subsection{Effect on cosmological parameter estimation}
\label{sub:}

It is clear from Figure~\ref{fig:Pkappa} that the change in the power
spectrum is large in the case of the AGN and DBLIMFV1618 scenarios.
The modification is, however, scale-dependent, which may help to ameliorate the
problem, since this cannot be reproduced by varying cosmological parameters which
predominantly affect the overall amplitude of the weak lensing power
spectrum. In other words, it might be possible to separate the effects
of baryonic feedback, or at least to identify them: the inferred values
for cosmological parameters from weak lensing statistics are
scale-dependent for the AGN and DBLIMFV1618 scenarios. 

We first investigate the effect  on the recovered value of $\sigma_8$, the rms fluctuation of matter in spheres of size $8\, h^{-1}\,$Mpc. A complication to our analysis is the limited accuracy of the
prescriptions for the non-linear power spectrum, be it Peacock \&
Dodds (1996) or the halofit approach (Smith et al.\ 2003) used here. We
therefore cannot predict $\xi_{+,\rm DMONLY}(\theta,z_{s})$ directly,
but the procedure outlined below is accurate as the predictions should
have the correct scaling as a function of $\sigma_8$. For the various
feedback models we first define the ratio

\be R_{+,\rm hydro}(\theta,z_{s})=\frac{\xi_{+,{\rm
      hydro}}(\theta,z_{s})}{\xi_{+,\rm DMONLY}(\theta,z_{s})},
\ee

\noindent as a function of source redshift $z_{s}$ and angular scale
$\theta$. Here $\xi_{+,\rm{hydro}}(\theta,z_{s})$ is the
correlation function measured for REF, DBLIMFV1618 or AGN, whereas
$\xi_{+,\rm DMONLY}(\theta,z_{s})$ is the DMONLY correlation function.
We use the halofit prescription (Smith et al.\ 2003) to compute
$\xi_{+,\rm halofit}(\theta, z_s; \sigma_8)$, keeping all other
cosmological parameters fixed to the reference values. We define the
ratio

\be R_{+,\rm halofit}(\theta,z_{s};\sigma_8)=\frac{\xi_{+,{\rm
      halofit}}(\theta,z_{s};\sigma_8)}{\xi_{+,{\rm ref, halofit}}(\theta,z_{s})}, 
\ee

\noindent where $\xi_{+,{\rm ref, halofit}}$ is the halofit prediction for the
reference cosmology. For each $z_{s}$ and $\theta$ we find the value
of $\sigma_8$ for which $R_{+,\rm halofit}(\theta,z_{s})=R_{+,\rm
  hydro}(\theta,z_{s})$. Hence, we compute by how much the value of
$\sigma_8$ needs to change from the reference value if one ignores
feedback processes and instead interprets the measurement of
$\xi_+(\theta,z_s)$ in a dark matter only framework. 

The top panels of Figure \ref{fig:delta_w} show the resulting value
$\Delta\sigma_8=\sigma_8-\sigma_{8,\rm ref} $ for which $R_{\rm halofit}=
R_{\rm hydro}$ as a function of $z_{s}$ for various values of
$\theta$. As we anticipated, the inferred values of $\sigma_8$ depend
on angular scale, but the effect of baryon physics is  modest,
 particularly for the REF simulation. For the AGN and  DBLIMFV1618 models  the effect is qualitatively the same and it is still only a few percent.

In models with a constant dark energy equation of state $P=w_0\rho$,
the change of $w_0$ mainly leads to a change in the amplitude of the
weak lensing power spectrum. It is therefore interesting to repeat the
same analysis for models with dark energy. Note that if we vary the value
of $w_0$, both the expansion history and the history of structure formation
change. Because we normalise  the amplitude of the fluctuations at the
present time, decreasing the value of $w_0$ results in fluctuations
that are larger at earlier times. On the other hand, a very negative
equation of state increases the expansion rate. Thus, the overall
amplitude of the two-point shear statistics at a given scale and
redshift depends on which of the two competing effects is dominant.

The bottom panels of Figure \ref{fig:delta_w} show the values $\Delta
w_0=w_0-w_{0,\rm ref}$ as function of the angular scale and source redshift.
Compared to the bias in $\sigma_8$, the change in the inferred value
of $w_0$ is more dramatic, reaching $20\%$ for
the AGN scenario, $10\%$ for the  DBLIMFV1618 scenario, and even for the REF scenario $w_0$  is few percent too
high. Note that we did not consider a redshift dependent equation of
state $w(z)$; also in that case we expect the estimated value of $w$
to change as a function of the angular scale, leading to similar
conclusions.  These findings suggest that baryon feedback can lead to
significant biases in cosmological parameters in the case of 
future cosmic shear studies. In the next section we will quantify
this in more detail.

\begin{figure*}
\begin{tabular}{|@{}l@{}|@{}l@{}||@{}l@{}|} 	
\psfig{figure=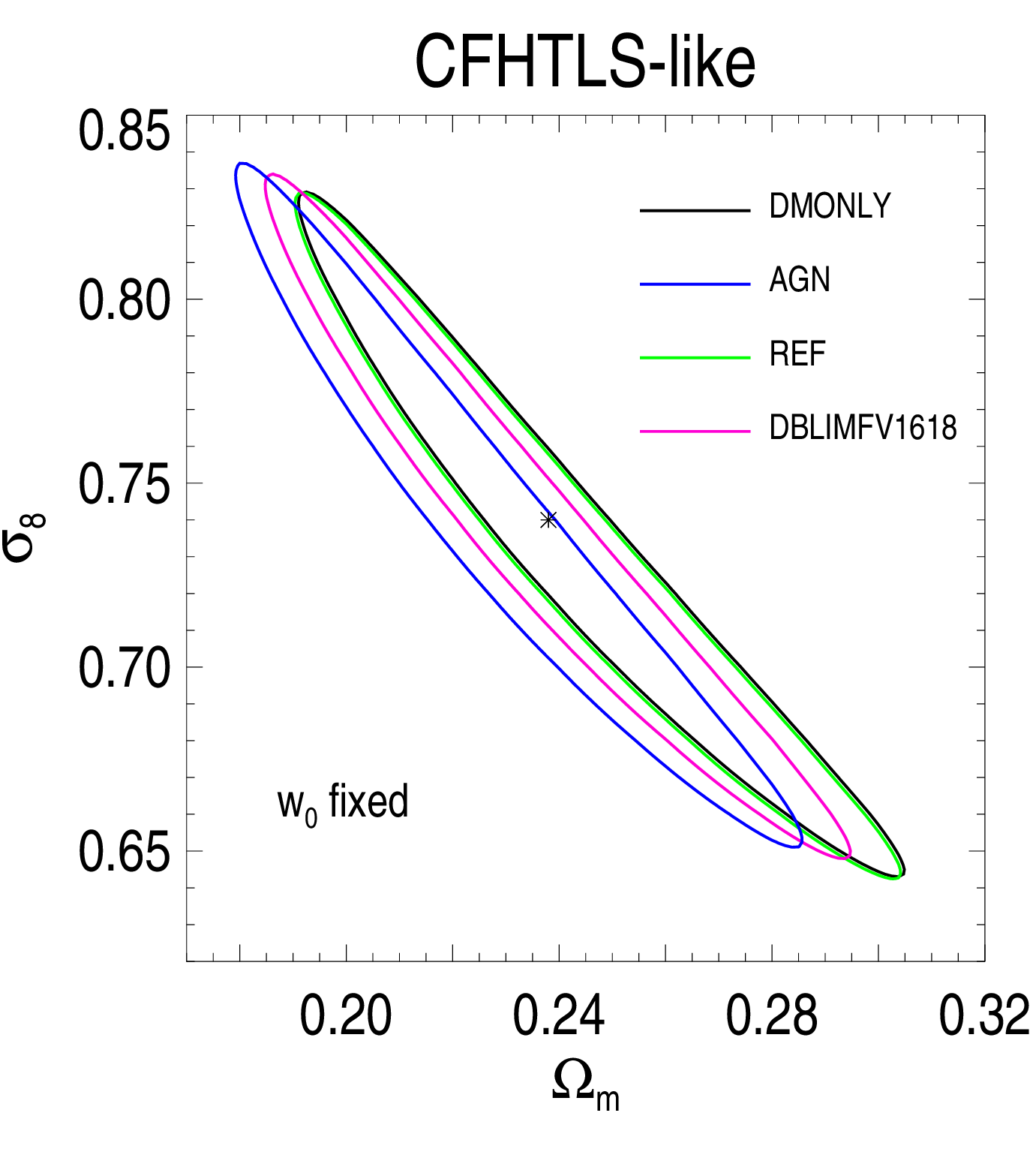,width=.33\textwidth}&\psfig{figure=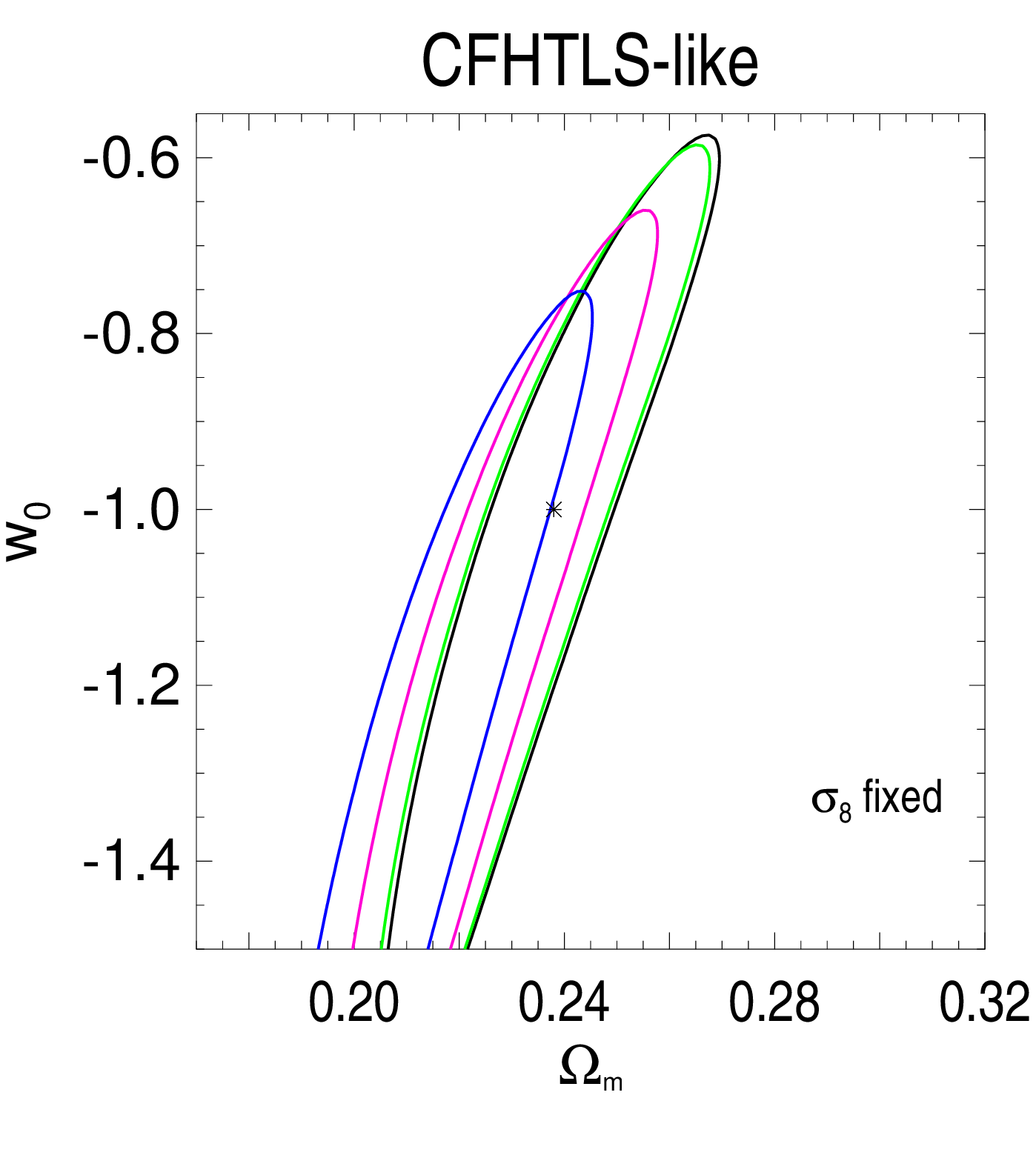,width=.33\textwidth}&\psfig{figure=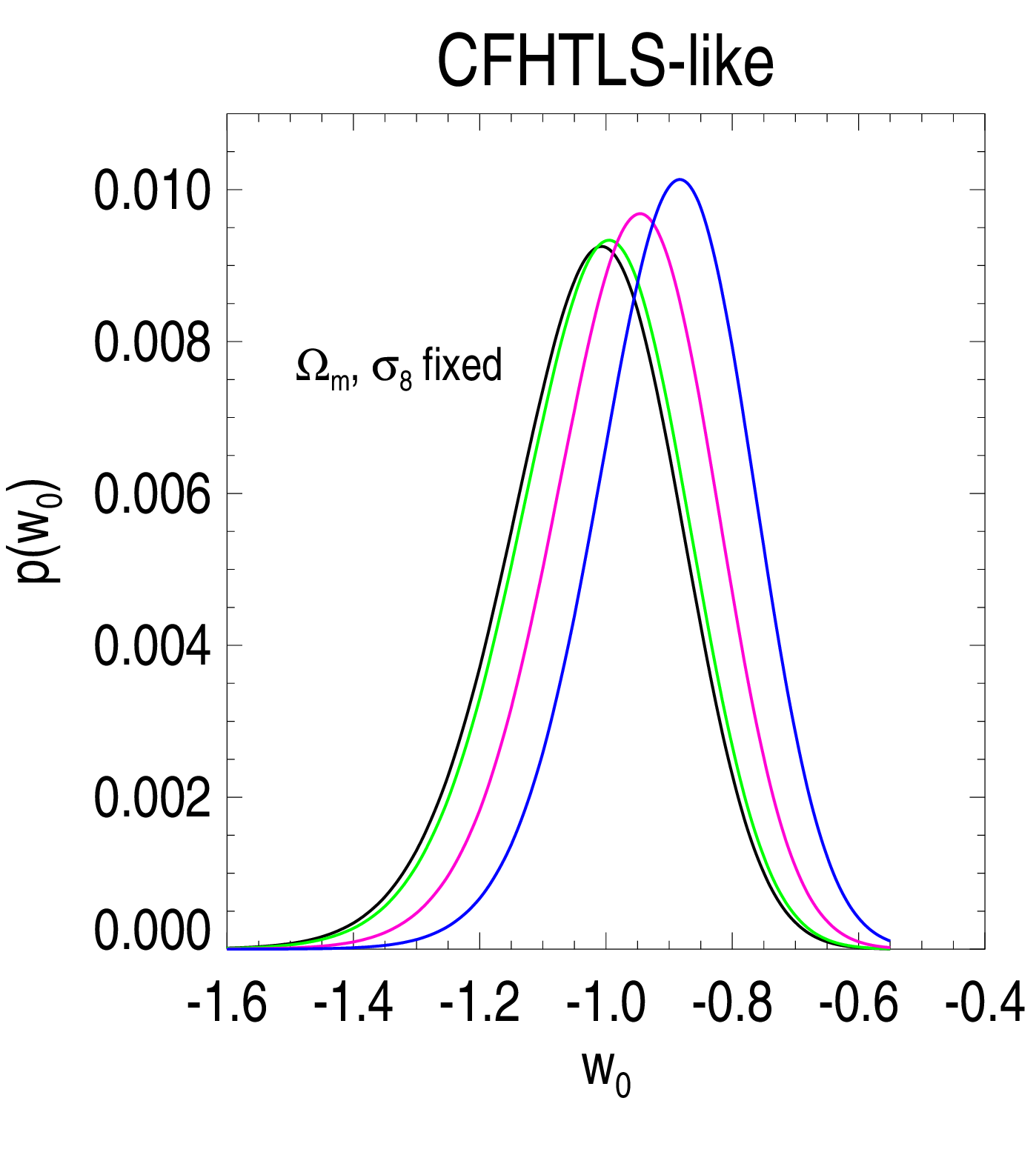,width=.33\textwidth}\\
\end{tabular}
\caption{\label{fig:like_CFHTLS} Left panel: $\Omega_{\rm m}-\sigma_8$
  likelihood contours for a CFHTLS-like survey.  Middle panel: $\Omega_{\rm m}-w_0$ likelihood contours for
  a CFHTLS-like survey. Right panel: posterior probability
  distribution for $p(w_0)$ obtained for $\sigma_8=0.74$ and
  $\Omega_{\rm m}=0.238$. Solid 
   contours mark the $68\%$ confidence regions. The shifts relative to the DMONLY case indicate the presence of bias due to baryonic effects. The biases are largest for the AGN scenario.}
\end{figure*}  

\section{Effect on likelihood Results} \label{sec:results}
\begin{figure*}
\begin{tabular}{|@{}l@{}|@{}l@{}||@{}l@{}|} 	
\psfig{figure=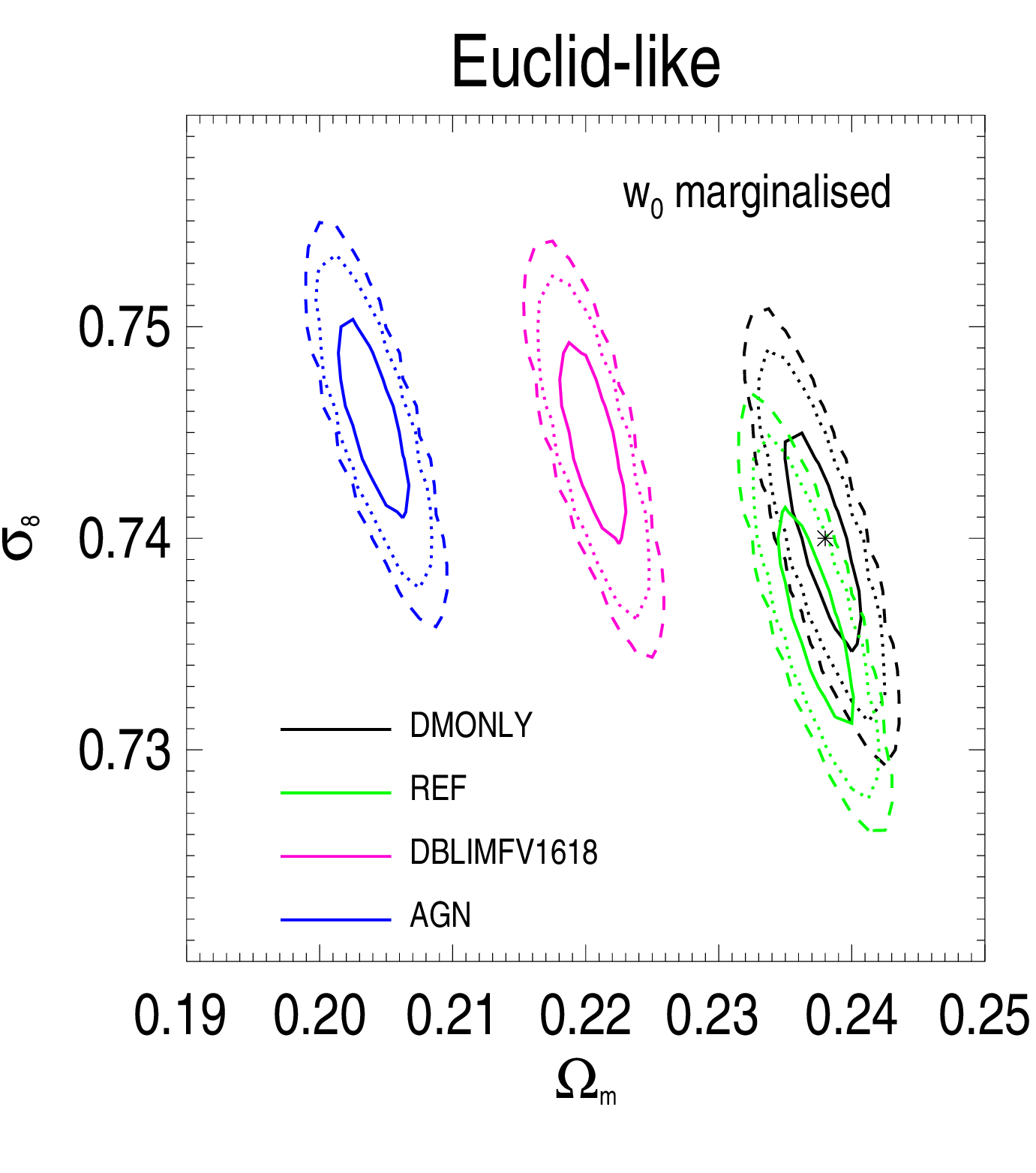,width=.33\textwidth}&\psfig{figure=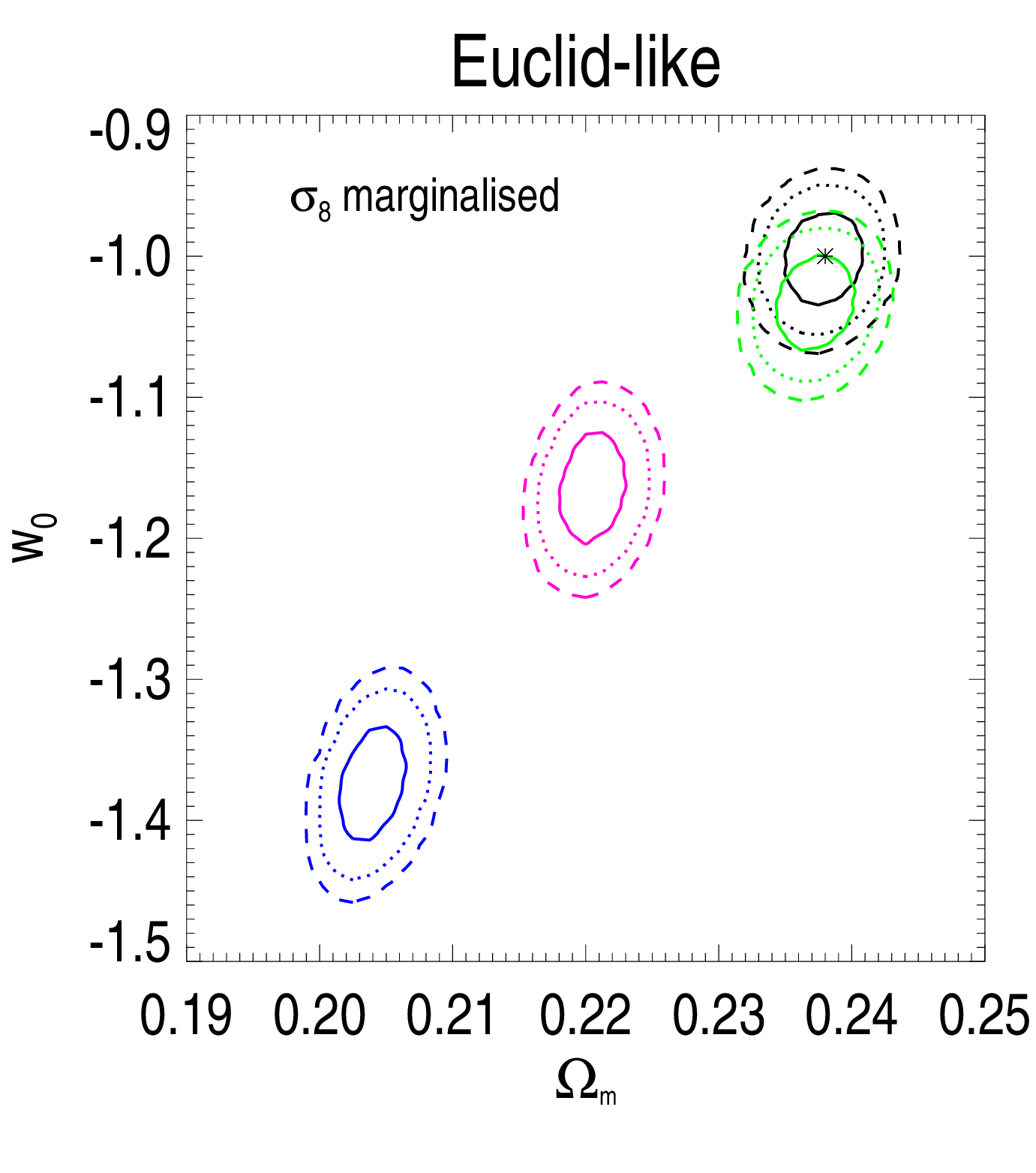,width=.33\textwidth}&\psfig{figure=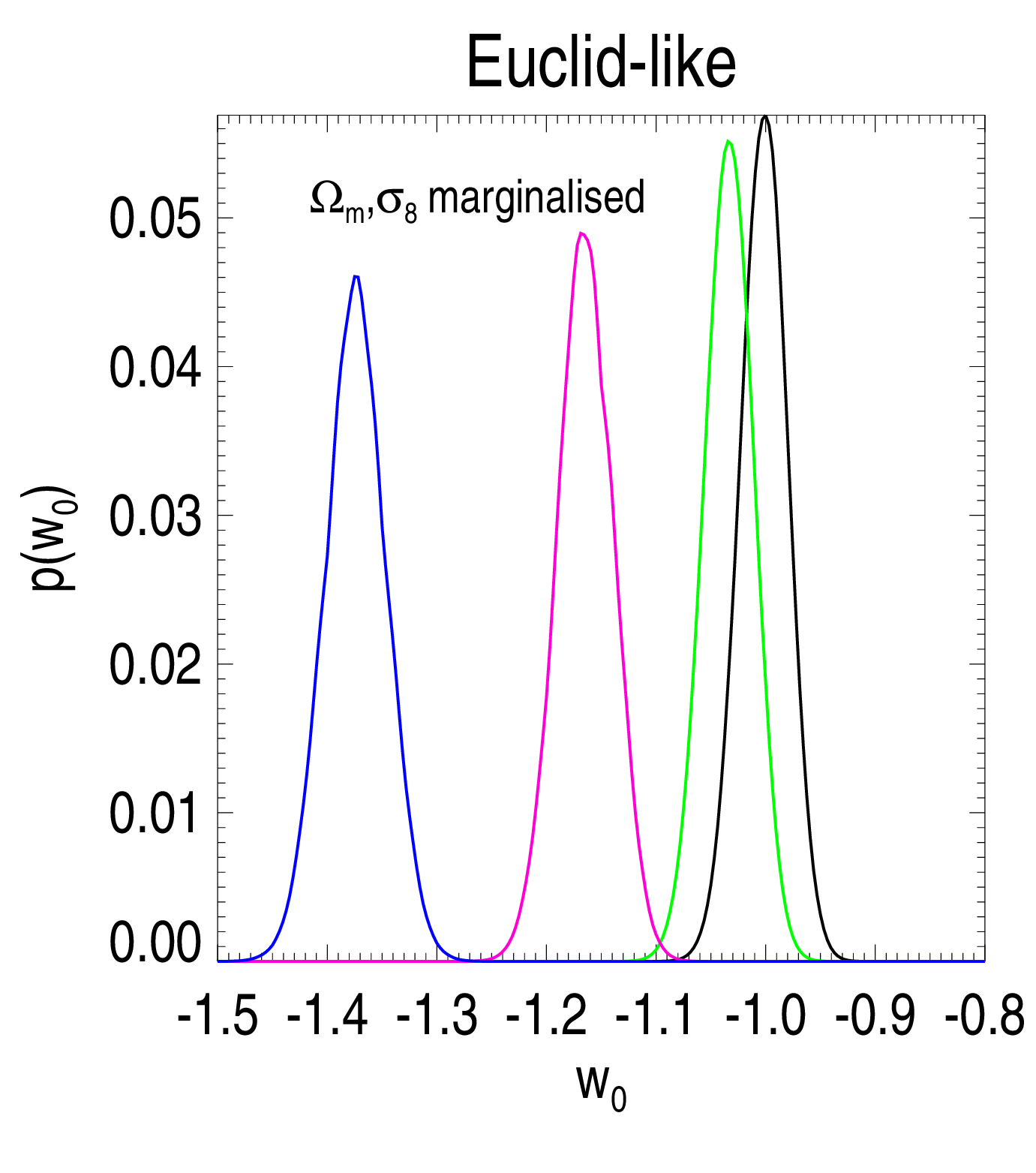,width=.33\textwidth}\\
\end{tabular}
\caption{\label{fig:like_Euclid} Left panel: joint constraints on
  $\sigma_8-\Omega_{\rm m}$ for a Euclid-like survey. Solid, dotted and
  dashed lines mark the $68\%$, $95\%$ and $99\%$ confidence regions
  respectively. Middle panel: joint constraints on $\Omega_{\rm m}-w_0$.
  Right panel: posterior probability distribution for $p(w_0)$
  marginalised over $\sigma_8$ and $\Omega_{\rm m}$.The shifts relative to the DMONLY case indicate the presence of bias due to baryonic effects. The biases are largest for the AGN scenario. Note that for the right panel the sign of 
  the shift differs from what was found in Figure \ref{fig:like_CFHTLS}, because we 
  now marginalise over the other parameters, rather than keeping them
  fixed.}
\end{figure*}  
In this section we quantify the bias in the inferred values for $w_0$
and $\sigma_8$ if one interprets weak lensing measurements using
dark matter  derived models. To do so, we perform a likelihood
analysis, where we define the posterior probability distribution as:

\begin{equation}
P( {\bf p}| {\bf d}) \propto P({\bf p}) {\mathcal L ({\bf d}| {\bf p})} \,
\end{equation}
\noindent where $ {\bf p}$ and ${\bf d}$ are vectors of parameters
and observed data, respectively, and the likelihood ${\mathcal L}
({\bf d}| {\bf p}) $ is given by:

\begin{equation}
{ \mathcal L} ({\bf d}| {\bf p}) \propto \exp{[ -1/2({\bf m(p)} -{\bf
      d}) {\mathcal C}^{-1} ({\bf m(p)} -{\bf d}) ^T ]}\,.
\end{equation}

\noindent Here ${\bf m(p)}$ is a model and ${\mathcal C}$ is the
covariance matrix. We chose the data vector ${ \bf d}$ to be
$P_\kappa(s)$ sampled at twenty scales between $s=10$ and $s=6000$ for
the six cross spectra introduced in section \ref{sec:2pt}. The
total data vector thus contains $120$ measurements. We compute
cosmological models $P_\kappa(s)$ at the same scales using the halofit
prescription for the non-linear power spectrum.  All the models use the transfer function  by  Eisenstein \& Hu (1998) which accounts for the effect that baryons have on the power spectrum during the radiation-dominated epoch. In this way the linear power spectrum of our models is similar (Eisenstein \& Hu is just an approximation) to the one used to establish the initial conditions of the OWLS simulations.

Throughout the paper we use flat priors  $\Omega_{\rm m}=[0.160,0.316]$
and $\sigma_8=[0.65,0.83]$, corresponding to the $\pm 3\sigma$
error-bars for WMAP7. We also adopt a flat prior of $w_0=[-2.00,
  -0.6]$. We emphasise  that uninformative priors are needed to study 
the bias, because (unbiased) information from  external data may
force the recovered values towards their unbiased values.

Following Takada \& Jain (2004), we compute the covariance matrix in
the Gaussian approximation. Note that the Gaussian approximation
breaks down at small scales. This is because  the non-linear evolution
of the density field causes the modes to mix, resulting in non-zero off-diagonal
terms. As a consequence, the value of the variance is
increased (Semboloni et al.\ 2007; Takada \& Jain 2009; Pielorz et al.\ 2010). Since we do not want to overestimate the impact of
baryon physics by underestimating the error bars at small scales, we only consider modes $s \leq 6000$, which
correspond to angular scales larger than $\theta \approx  3.5{\rm arcmin}$.

As before, we do not use the data vector ${\bf d}_{\rm hydro}$,
which is measured from the simulations. Instead we construct
a new data vector ${\bf d}_{\rm fit}$:
\be
\label{eq:calibrate} {\bf d}_{\rm fit}= \frac{{\bf d}_{\rm halofit}}{{\bf d}_{\rm DMONLY}}\times {\bf d}_{\rm hydro},
\ee

\noindent where ${\bf d}_{\rm hydro}$ corresponds to $P_\kappa(s)$
from the REF, AGN, or DBLIMFV1618 simulation, and where ${\bf d}_{\rm
  DMONLY}$ is computed using the DMONLY simulation. The halofit
approach is used to calculate ${\bf d}_{\rm halofit}$, which
quantifies the dependence on the cosmology.  If the  halofit model  described the power spectrum  of the DMONLY simulation  perfectly then the ratio in Eq. \ref{eq:calibrate} would be unity  for any component of the data vector and ${\bf d}_{\rm fit}$ would be merely ${\bf d}_{\rm hydro}$. This is, however, not the case because the halofit model has limited accuracy; moreover the set of simulations we use is relatively small so the measured power spectrum is affected by sampling variance. We are only interested in the comparison between dark matter only simulations and  simulations with baryonic feedback. By using Equation \ref{eq:calibrate} we can compute the bias in cosmological parameters that one would obtain by neglecting the existence of baryons, while minimising the limitations of the halofit prescription and sampling variance due to the finite size of the simulations. 

We first consider a survey with area $A=200~ {\rm deg}^2$, galaxy  number density $n=15~ {\rm
  gal/arcmin^2}$ and  intrinsic ellipticity dispersion of galaxies $\sigma_e=0.44$, which roughly corresponds to the
CFHTLS-Wide survey (e.g., Hoekstra et al.\ 2006; Fu et al.\ 2008).  The
joint constraints on $\Omega_{\rm m}$ and $\sigma_8$ are presented in the
left panel of Figure~\ref{fig:like_CFHTLS}, whereas the constraints on
$\Omega_{\rm m}$ and $w_0$ are displayed in the middle panel. Note that we
keep the other parameter(s) fixed, rather than marginalising over
them, because the statistical power of the CFHTLS is too limited to
constrain them.  The shifts in the likelihoods, compared to the DMONLY
contours, indicate that the cosmological parameters are biased if a
dark matter power spectrum is used to interpret the data. In agreement
with our earlier findings, the shifts are small for the REF model and
larger for the other two scenarios.

Based on the results of the previous section, we are most concerned
about a bias in $w_0$. The right panel of Figure~\ref{fig:like_CFHTLS}
shows the posterior probability for $w_0$, when the other parameters
are kept fixed at their fiducial values. This result suggests that a dark matter only modelling of the CFHTLS signal will
likely predict a value of $w_0$ which is about $10\%$ less negative  than the
real one. However, the DMONLY peak lies at the edge of the $68\%$
confidence region of the AGN scenario, thus the bias is similar to the
statistical uncertainty.  Note that since the CFHTLS-Wide has limited
statistical power due to its modest survey area, we fixed the values
of $\sigma_8$ and $\Omega_{\rm m}$ to constrain $w_0$, which affects the
bias. We will see below that the bias is different (and increases) if we use a flat
prior.

To reach a precision of a few percent, much larger projects are being
planned.  Of these, we focus on the space-based Euclid mission
\cite{Reetal10}, which aims to survey an area $A=20000~ {\rm deg}^2$
with a number density of galaxies of $n=30~ {\rm gal/arcmin^2}$ and
intrinsic ellipticity dispersion $\sigma_e=0.33$. This leads to much
smaller statistical errors on the cosmological parameters. As shown in
Figure \ref{fig:like_Euclid}, in this case the effect of baryons leads
to significantly biased results. The left panel shows the constraints
on $\Omega_{\rm m}$ and $\sigma_8$, while marginalising over the value of
$w_0$. The likelihood contours for both the AGN and DBLIMFV1618 models
are shifted towards lower values for $\Omega_{\rm m}$, and an only slightly
higher $\sigma_8$. The consequences for the constraints on $w_0$ are
even more dramatic, as is clear from the middle panel of
Fig. \ref{fig:like_Euclid}. The bias in $w_0$ reaches almost 40\% for
the AGN scenario, as is also evident from the marginalised probability
distribution shown in the right panel. The large bias is the
consequence of the large modification of the power spectrum at
intermediate scales.  Note that the shift in the value of $w_0$ in the
right panel is in the direction opposite to what was found in
Figure~\ref{fig:delta_w}, because here we marginalised over the other
parameters whereas they were kept fixed in Figures~\ref{fig:delta_w}
and~\ref{fig:like_CFHTLS}.

\begin{figure}
\psfig{figure=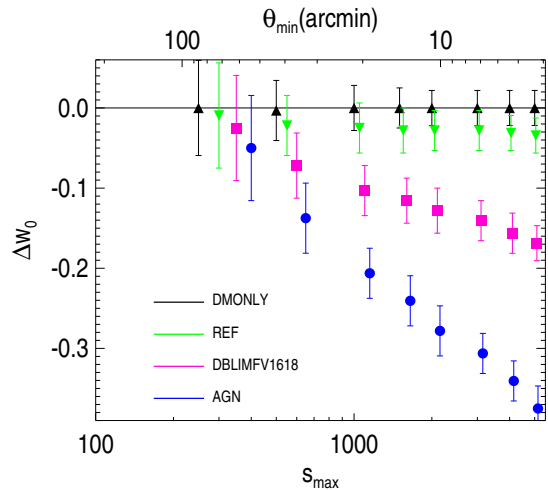,width=.45\textwidth}
\caption{\label{fig:k_max} Difference $\Delta w_0$ between the best fit
  value of $w_0$ and the true reference value $w_0=-1$ as a function of the
  maximum angular wavenumber $s_{\rm max}$ (or minimum scale $\theta_{\rm
    min}$) that is included in the likelihood analysis. The error bars
  represent the resulting  $1\sigma$ uncertainties on $w_0$. Avoiding high wave numbers allows one to reduce the bias
  affecting the cosmological parameters at the cost of an increase of the statistical errors.}
\end{figure}  

Since the bias arises due to differences in the power spectrum at
relatively small angular scales $\theta$, or large $s$, it is
interesting to examine whether the bias can be reduced by leaving the
large wave numbers out of the likelihood analysis. Figure
\ref{fig:k_max} shows the maximum likelihood value of $w_0$ as a
function of the maximum angular wave number $s_{\rm max}$, while
marginalising over $\Omega_{\rm m}$ and $\sigma_8$. The error bars on the
points indicate the 68\% confidence regions. The bias is no longer
statistically significant if the posterior probability for $w_0$ peaks
well within one sigma from the reference value $w_0=-1$. For the AGN
model this is only achieved when $s < 500$ (this corresponds to a real
space separation of more than 40 arcminutes!).  However,
drastically limiting the range of scales increases the statistical
uncertainty by almost a factor of three. We therefore conclude that
this approach is not viable and that one needs to account for the
effects of baryon physics when computing the constraints on
cosmological parameters.

\section{Reducing the bias using a simple model}
\label{sec:toy_model}

The results presented in the previous section suggest that one cannot
ignore the effects of baryon physics on the matter power spectrum in
the case of future lensing surveys. A complication is that the bias
itself depends strongly on the details of the feedback model, but that
we do not know for sure which of the feedback scenarios (and
parameters) is correct. However, baryon physics also has an impact on
other observables, which can be used to discriminate between models.

For example, McCarthy et al.\ (2010) showed that the AGN and REF
simulations yield haloes with significantly different gas
fractions. Similarly, the amount of gas that cools to form stars is
different, leading to different luminosities.  For both observables,
the AGN simulation provided a good match to observations of groups of
galaxies whereas the REF simulation did not. In principle, such
observations can be used to select hydrodynamic simulations that best
describe our Universe. In this section, we will explore a different
approach and show that those same observables can be used to modify
the dark matter power spectrum such that it accounts for most of the
effects of baryon physics.

\subsection{Halo model}
To predict the matter power spectrum analytically, we take advantage
of the fact that the clustering of haloes of a given mass is known in
the linear regime and that the average density profiles of dark matter
haloes are specified by their mass. As shown by Seljak \ (2000), this
``halo model'' approach can reproduce the power spectrum into the
non-linear regime, although some parameters have to be calibrated
using numerical simulations. 

The power spectrum is computed as the sum of two terms. The 
first one describes the correlation of the density fluctuations
within the same halo. This Poisson term $P^{\rm P}(k)$ dominates
on small scales and is given by

\begin{equation} 
P^{\rm P}(k)=\frac{1}{(2 \pi)^3} \int {\rm d}\nu f(\nu)
\frac{M(\nu)}{\bar\rho} y[k,M(\nu)]^2,
\end{equation}

\noindent where $\bar\rho$ is the mean matter density and $y[k,M(\nu)]$ is
the Fourier transform of the  density profile of a halo with
virial mass $M(\nu)$ normalised such that:

\be
y[k,M]=\frac{\int_0^{r_{\rm vir}} 4 \pi r^2 {\rm d}r \frac{\sin(kr)}{kr} \rho(r)}{\int_0^{r_{\rm vir}} 4 \pi r^2 {\rm d} r \rho(r)}\,,
\ee

\noindent  where $\rho(r)$ is the density profile of the halo and $r_{\rm vir}$ its virial radius. The peak height $\nu$ of such an overdensity is defined as
\begin{equation}
\nu=\left[\frac{\delta_{\rm c}(z)}{\sigma(M)}\right]^2,
\end{equation}

\noindent where $\delta_{\rm c}$ is the linear theory value of a spherical
overdensity  which  collapses at a redshift $z$. $\sigma(M)$ is
the rms fluctuation in spheres that contain mass $M$ at an initial
time, extrapolated to $z$ using linear theory. We use \cite{ShTo99}:

\begin{equation}
\nu f(\nu)=A (1+\nu^{\prime-p}) \nu^{\prime 1/2} \exp({-\nu^\prime/2}),
\end{equation}

\noindent where $\nu^\prime=a\nu$ with $a=0.707$ and $p=0.3$. The
normalisation constant $A$ is determined by imposing $\int f(\nu) {\rm d}\nu
=1$. Note that the function $f(\nu)$ is related to the halo mass function ${\rm
  d}n/{\rm d}M$ through

\begin{equation}
\frac{{\rm d}n}{{\rm d}M}{\rm d}M=\frac{\bar\rho}{M}f(\nu){\rm d}\nu.
\end{equation}

\noindent The second term, $P^{\rm hh}(k)$, describes the clustering of haloes
and dominates on large scales. It is given by

\begin{equation}
 P^{\rm hh}(k)=P_{\rm lin}(k) \Big ( \int {\rm d}\nu f(\nu) b(\nu) y [k,M(\nu)]
\Big)^2,
\end{equation} 

\noindent where $P_{\rm lin}(k)$ is the linear power spectrum, and
the halo bias $b(\nu)$ is given by (Mandelbaum et al. 2005):

\begin{equation}
b(\nu)=1+\frac{\nu^\prime-1}{\delta_c}+\frac{2p}{\delta_c(1+\nu^{\prime
    p})},
\end{equation}

\noindent with $a=0.73$ and $p=0.15$. Finally, one generally assumes that the
density profile is of the form \cite{Naetal95}:

\begin{equation}
\rho(r)\propto\frac{1}{r(r+r_s)^2},
\end{equation}

\noindent where $r_s$ is the scale radius. Numerical cold dark matter
simulations have shown that this NFW profile is a fair description of
the radial matter distribution for haloes with a wide range in mass.
They also indicate that $r_s$ is not a free parameter, but that it is
related to the virial mass (albeit with considerable scatter). It is
customary to account for this correlation by specifying a relation,
between the concentration $c=r_{\rm vir}/r_s$ and the virial mass. We
use the mass-concentration relation derived by Duffy et al. (2008).  The
virial mass and radius are related through $M_{\rm vir}=(4\pi/3)
r_{\rm vir}^3{\rho_c}\delta_{\rm vir}$, where we use the fitting
formula of Bryan \& Norman (1998) to compute $\delta_{\rm vir}(z)$.

It is typically assumed that the total matter density can be described
by the NFW profile. However, if the stars and the gas do not follow
the dark matter profile, then the resulting mass profile and thus the power
spectrum will be different. In the remainder of this section we will
explore whether it is possible to simply modify the density profile to
better describe the distribution of the baryons, and whether this
model can reduce the biases discussed in the previous section.  Because of the back-reaction of the baryons on the dark
matter, baryonic effects will also induce changes in the dark matter
distribution within haloes (e.g., Duffy et al. 2010) and in the dark
matter power spectrum (van Daalen et al. 2011). For simplicity, we
will, however, ignore this complication.

\subsection{An improved halo model}

The halo model is only a good description of the total matter
distribution if the baryons trace the underlying dark matter
distribution. 

 As feedback processes redistribute the baryons (which
 make up $\sim 17\%$ of the total amount of matter), we
expect the total matter power spectrum to be modified. In this
section we explore whether it is possible to predict the power
spectrum from hydrodynamic simulations using `observations' of the
gas fraction and stellar mass (or luminosity). 

 Somogyi \& Smith (2009) showed that in order to construct accurate  models that include CDM and  baryons  one should treat the density field as a two-component fluid. The procedure that we use here, which is to compute the linear matter power spectrum using the   Eisenstein \& Hu (1998) transfer function, corresponds to a one-component fluid approximation.  Somogyi \& Smith (2009) showed  this approximation leads to a biased  power spectrum, in particular, for high redshifts and for the baryonic component. However, following their studies, the bias  is less than $1\%$ for $z\lesssim 3$. Since this effect is much smaller than the effect we are interested in, we can neglect  it.  Furthermore, we assume that any modification of the final power spectrum is caused by phenomena which affect the baryonic component  in haloes  and happen  after the halo has collapsed. This  implies that haloes collapse in the same way as in a CDM-only Universe. We know this assumption is not completely  correct, as van Daalen et al. 
(2011) have demonstrated that the back reaction of the baryons onto the 
CDM causes the power spectrum of the CDM component to differ from that 
of a dark matter only simulation. However, the effect of the back 
reaction on the total matter power spectrum is smaller than the effect caused by the change of the 
distribution of baryons. 

We start by modifying the density profile to better describe the
distribution of the stars and the gas. Note that the NFW profile with the mass-concentration relations of Duffy et al. (2008)  is
still used to describe the dark matter distribution. The stellar mass
in a halo is much more concentrated  than either the gas and dark matter
components, and we therefore approximate its distribution by a point
mass. To describe the gas component, we use a single
$\beta$-model, which provides a fair description to X-ray observations
in groups and clusters of galaxies (e.g., Cavaliere \& Fusco-Femiano 1976;
Reiprich \& B{\"o}hringer 2002; Osmond \& Ponmann 2004).  The corresponding density profile $\rho_{\rm gas}$, is
given by
\be
\rho_{\rm gas}(r)=\rho_0\Big[ 1+\Big(\frac{r}{\alpha
    r_{500}}\Big)^2\Big]^{-3\beta/2}, \label{eq:betamod}
\ee 
\noindent where $\alpha$ is the ratio between the characteristic scale
of the gas profile (i.e. the core radius) and $r_{500}$, the radius of a sphere with average density $500$ times the critical density. The value for the slope $\beta$ is usually
considered a free parameter in the fit to X-ray data, but we note
that for a hydrostatic isothermal sphere its value corresponds to the
ratio of the specific energy in galaxies to the specific energy in the
hot gas (e.g.,\ King 1972; Jones \& Foreman 1984; see also  Mulchaey 2000 for a review).  Finally, note that although different models have been used in the literature  (e.g.,\ Osmond \& Ponmann 2004; Arnaud et al.\ 2010),  the differences between these models are too small  to be important here.

%

\begin{figure*}
\begin{tabular}{cc}
\psfig{figure=mccarthy10.1.ps,width=.45\textwidth}&\psfig{figure=mccarthy10.2.ps,width=.45\textwidth}
\end{tabular}
\caption{\label{fig:mccarthy} Left panel: relation between $M_{500}$
  and gas fraction in haloes for  REF, DBLIMFV1618  and AGN. Right panel: relation between $M_{500}$ and the
 stellar  luminosity in the $K$-band $L_{K s}$ of the brightest galaxy in the halo. The solid lines
  represent the best linear fits to the points (see text). The simulation data was taken from McCarthy et al. (2010), who showed that the AGN model agrees with the available observations of groups. }
\end{figure*}  

Our model provides a convenient description of the final mass
distribution  because even though the parameters depend on the details of the
various feedback processes, they can be constrained observationally. The most important observable is the gas fraction as a function of halo mass: the gas has an extended profile and the overall profile depends on the fraction of gas which is left in the halo. This has been measured by McCarthy et al.\ (2010; 2011) for the OWLS simulations and effectively determines the
value of $\rho_0$ in Eqn.~\ref{eq:betamod}. The results are presented in the left panel of
Figure~\ref{fig:mccarthy}; it is clear that different feedback models
lead to rather different scaling relations. 

\begin{figure*}
\psfig{figure=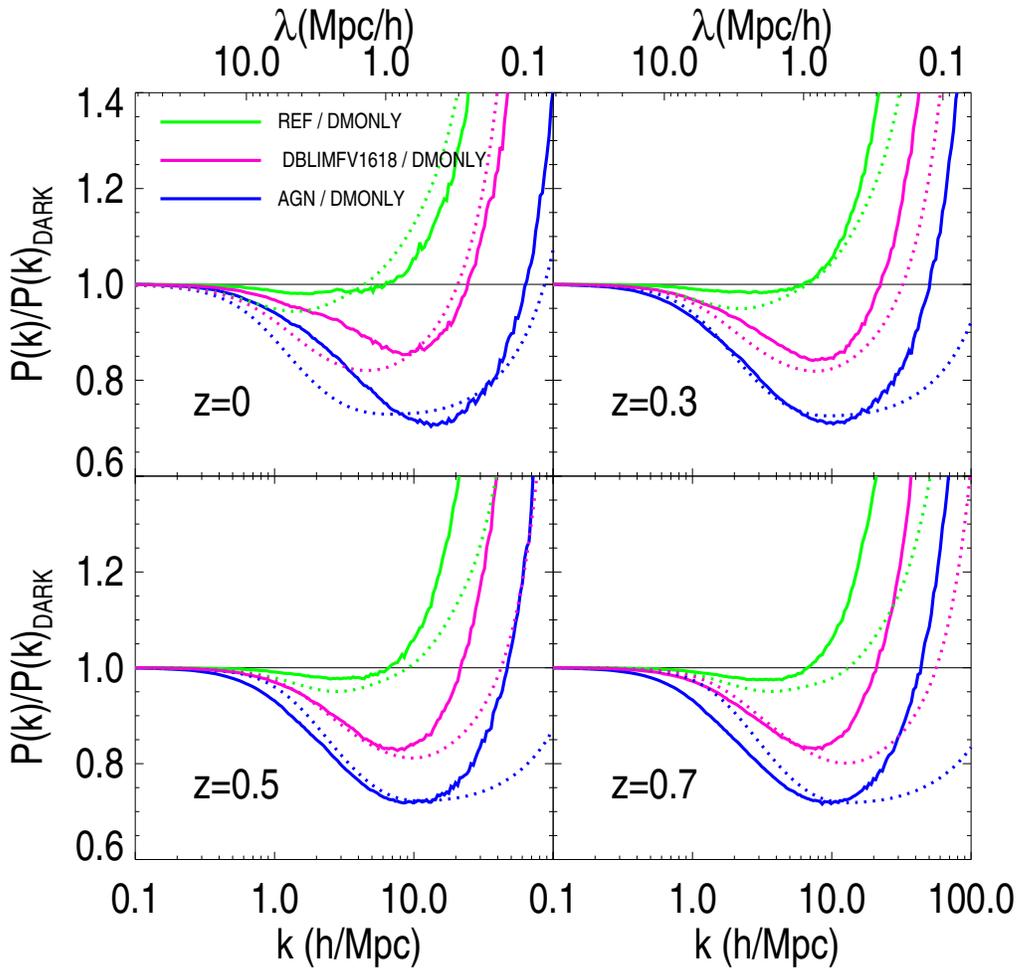,width=.90\textwidth}
\caption{\label{fig:toy_model} Comparison of the ratio of the power
  spectra REF/DMONLY, DBLIMFV1618/DMONLY and AGN/DMONLY measured from the simulations
  (solid lines) and predicted by  our simple halo model (dotted lines) which has been tuned to fit the gas and stellar mass fractions predicted in each scenario (see text). The model
  matches the simulations fairly well for $k<10~h~{\rm Mpc}^{-1}$.}
\end{figure*}

The values of $\alpha$ and $\beta$ have been measured from the
simulations.  We note that these values have a large dispersion and
depend on the mass; however, we take here an average value measured
over a range of masses $10^{13}\,h^{-1} M_\odot<M_{500}<10^{15}\,h^{-1} M_\odot $. This corresponds to $\alpha=0.01$ and
$\beta=0.4$ for the AGN and DBLIMFV1618 models and $\alpha=0.2$ and
$\beta=0.9$ for the REF model.  
  The power spectrum is mildly sensitive to the choice of the slope $\beta$ as this slope
defines how fast the gas power spectrum declines for large $k$. The
AGN model, for which a large fraction of the gas has been ejected
beyond $r_{500}$, has a power spectrum whose shape is less sensitive
to the choice of $\beta$.

The right panel shows the relation between $M_{500}$, the mass of a sphere  with average density  $500$ times the critical density of the universe,  and $L_{Ks,{\rm BCG}}$,
the stellar luminosity of the (central) brightest galaxy of the halo in the $K-$band as
predicted by McCarthy et al.\ (2010) using the metallicity-dependent population synthesis model of Bruzual \& Charlot (2003). 
To constrain the parameters of our extended halo model, we fit a power law to the data in  Figure \ref{fig:mccarthy} and we find:
\ba
f_{\rm gas}/f_{b}(r_{500})&=&0.15 \log_{10}(M_{500})-1.49 ~~~~ {\rm REF}\nonumber\\
f_{\rm gas}/f_{b}(r_{500})&=&0.30 \log_{10}(M_{500})-3.40 ~~~~ {\rm DBLIMFV1618}\nonumber\\
f_{\rm gas}/f_{b}(r_{500})&=&0.40 \log_{10}(M_{500})-4.94 ~~~~ {\rm AGN}\nonumber\\
\log_{10}(L_{Ks,{\rm BCG}})&=&0.62 \log_{10}(M_{500})+4.19  ~~~~ {\rm REF}\nonumber\\
\log_{10}(L_{Ks,{\rm BCG}})&=&0.81 \log_{10}(M_{500})+1.39  ~~~~ {\rm DBLIMFV1618}\nonumber\\
\log_{10}(L_{Ks,{\rm BCG}})&=&0.83 \log_{10}(M_{500})+0.23  ~~~~ {\rm AGN}\nonumber
\ea
To compute the stellar mass from the stellar luminosity, we measure  the average mass-to-light ratio of the BCGs directly from the simulations.
 We find $M_\star/L_\star=0.65 M_\odot/K_\odot $ for AGN, $0.32 M_\odot/K_\odot $ for DBLIMV1618, and $0.33 M_\odot/K_\odot $ for REF.  
We could have measured the stellar mass of the BCG directly from the simulations,
 but  since we want to use quantities which can  in principle be observed,
 we  derive the stellar mass using the same procedure  that one needs to use for real data.

To determine the fraction of baryons (stars + gas) that remains within
$r_{500}$, we use the scaling relations shown in Figure
\ref{fig:mccarthy}. We then assume that the gas fraction within $r_{\rm
  vir}$ is the same as within $r_{500}$, although we note that may not be correct.  The
resulting halo density profile is the sum of the stellar, gas and dark
matter profiles weighted by their respective fractions.  
We assume that the ejected gas is still associated with the the
parent halo and make ad-hoc assumption that the gas is distributed uniformly between
$r_{\rm vir}<r<2 r_{\rm vir}$. We add this uniform component to the function $y[k,M(\nu)]$ so that the total mass of the halo is unchanged. The density distribution of the gas at large radii can been derived,  using X-ray observations, only for  bright clusters  \cite{Reetal08}, and it  might be overestimated \cite{Sietal11}.  Stacked X-ray (e.g., Dai et al.\ 2010)  and SZ observations (e.g., Afshordi et al.\ 2007; Komatsu et al. 2011)  could  allow one to obtain a more accurate estimation of the gas distribution  even for less massive haloes. We will measure  the distribution of this gas using the simulations and evaluate the importance of its modelling in a future work. 

Under these assumptions our model has more power than the simulations
for high $k$. This is due to an overestimate of the stellar mass which
can be explained as follows.  Using the simulations data by McCarthy et
al.\ (2010) we can only establish a relation between $M_{500}$ and
$L_{Ks,{\rm BCG}}$ for haloes with $M_{500} >5 \times 10^{12} \, h^{-1}\,M_\odot 
$. Observationally the relation between the fraction of stars and halo
mass is approximately linear down to a certain halo mass; for haloes with mass below
a few times $10^{12}\, h^{-1}\, M_\odot$ this relation is poorly constrained. However,
observations suggests a stellar fraction for Milky Way size haloes
which is about $0.01$, suggesting that the stellar-mass to halo-mass
relation has to change.  This observed relation between luminosity and
halo mass agrees well (see e.g.\ Moster et al.\ 2010) with models
constructed using halo occupation distributions ( e.g.,\ Peacock \&
Smith 2000; Seljak 2000; White 2001; Berlind \& Weinberg 2002) and
conditional luminosity functions (Yang et al.\ 2003; van den Bosch et al.\ 2007; Cacciato et
al.\ 2009), as well as with semi-analytic models (e.g.\ Croton et
al.\ 2006; Bower et al.\ 2006). Those approaches all show that there
is a characteristic mass $M\approx 10^{12.5}\, h^{-1}\, M_\odot$ for which star
formation is particularly efficient, whereas for small masses the
fraction of stars in haloes decreases steeply. This should also happen
in our simulations although we lack the resolution required to measure
the stellar mass-to-light ratio for small haloes.  Thus, we just
assume that the mass of the stars is $0.01\,M_{500}$ for $M_{500} <
5\times 10^{12}\, h^{-1}\, M_\odot$ which approximately corresponds to the
fraction of stars in a Milky Way size halo. Note that while low-mass
haloes do not contribute much at small $k$, they will dominate the
signal if the stellar fraction is too large.

In Figure \ref{fig:toy_model} we compare the ratios of the power
spectra measured from the simulations to the ones predicted by the
model.  Despite its simplicity, the model is able to reproduce the
main features of the power spectrum for all scenarios for redshifts to
which the weak lensing signal is most sensitive.
Although it is a clear improvement over the pure dark matter model,
the model is still too simplistic  and does not  perform equally well at all redshifts and all scales.
The accuracy of the toy model can clearly be improved and we will aim
to do so in future work. In particular, we will examine requirements on
the types of observations that will be needed to constrain the model
and the mass and length scales that need to be studied.

\begin{figure}
\psfig{figure=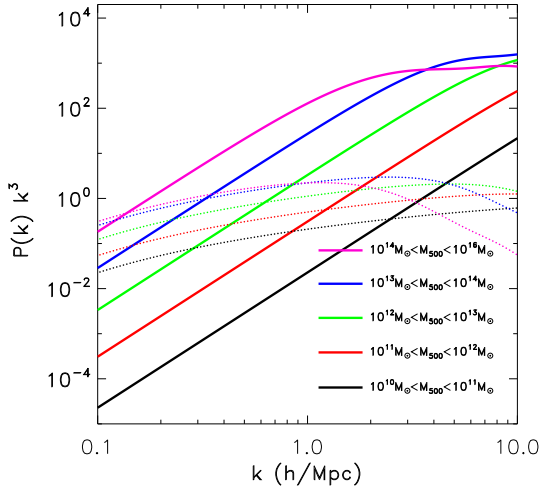,width=.45\textwidth}
\caption{\label{fig:halo_masses} Contribution of haloes of different
  masses to the one halo-term $P^{\rm P}(k)$ (solid lines)  redshift $z=0.5$. We show for comparison the  halo-halo term $P^{\rm hh}(k)$ (dotted lines)  for the same mass ranges. At scales where the baryonic feedback is important, the dominant contribution is the Poisson contribution of haloes with $M_{500}\gtrsim 10^{13}\, h^{-1}\, M_\odot$.}
\end{figure}

\begin{figure*}
\begin{tabular}{|@{}l@{}|@{}l@{}||@{}l@{}|}
\psfig{figure=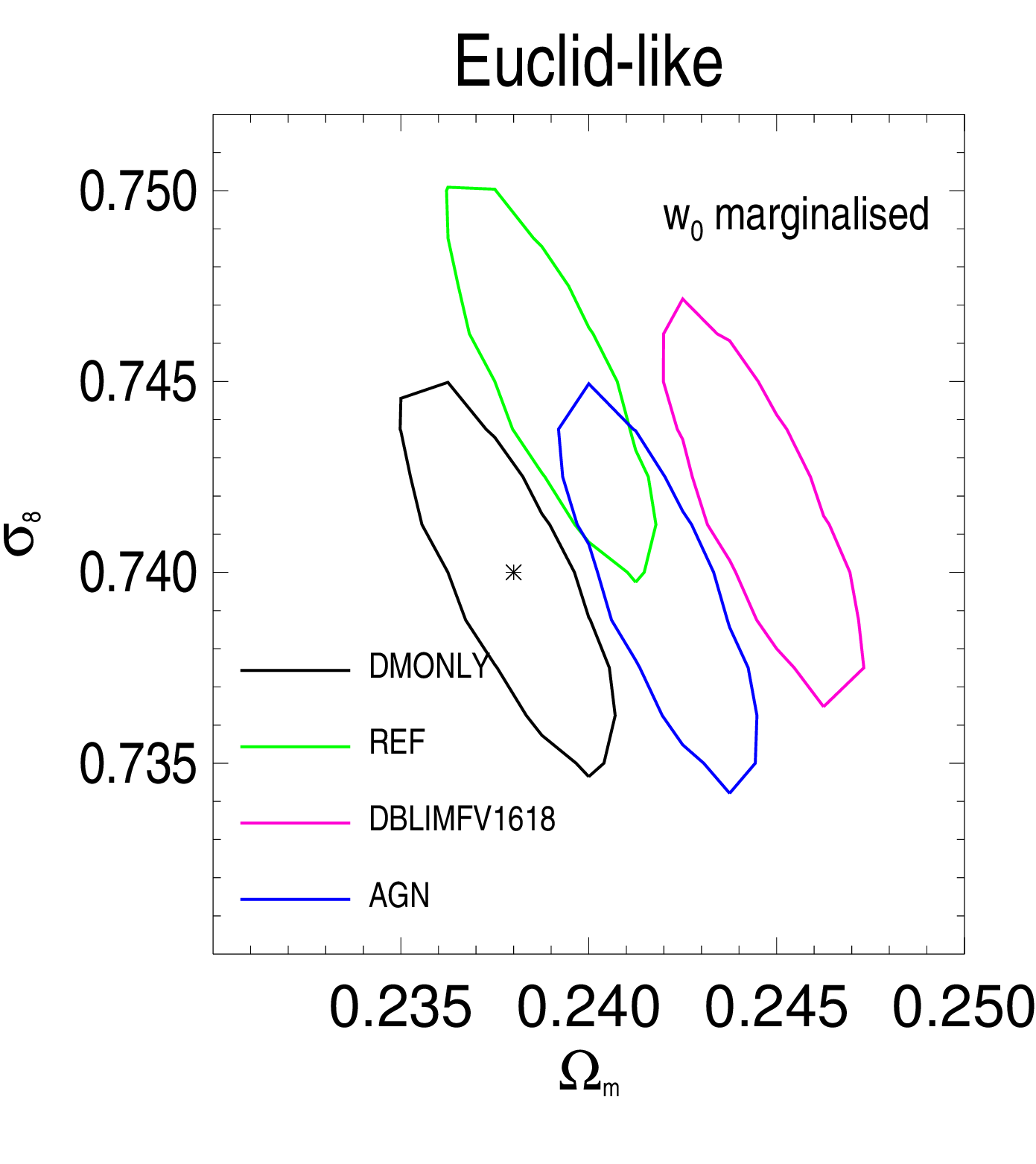,width=.33\textwidth}&\psfig{figure=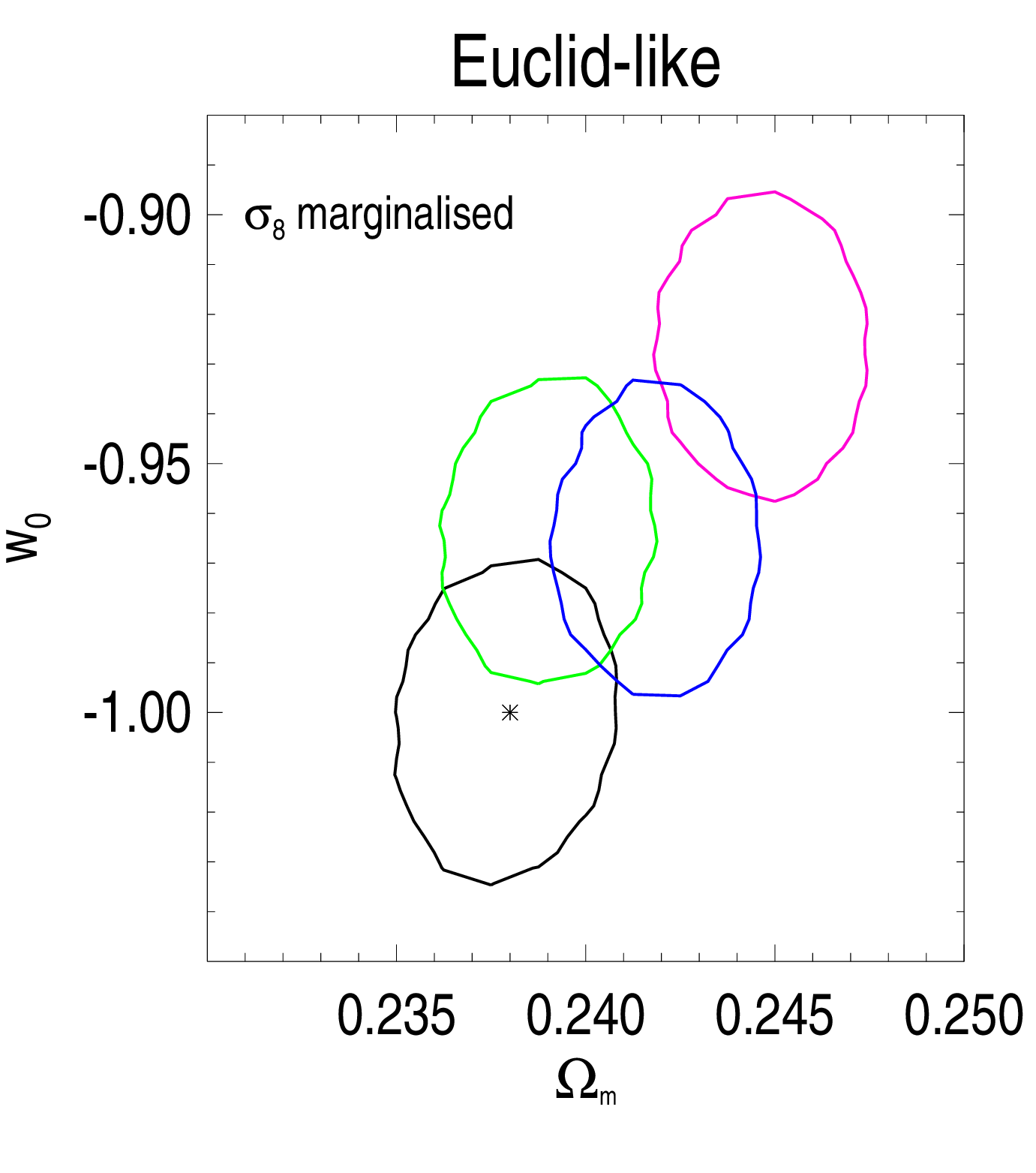,width=.33\textwidth}&\psfig{figure=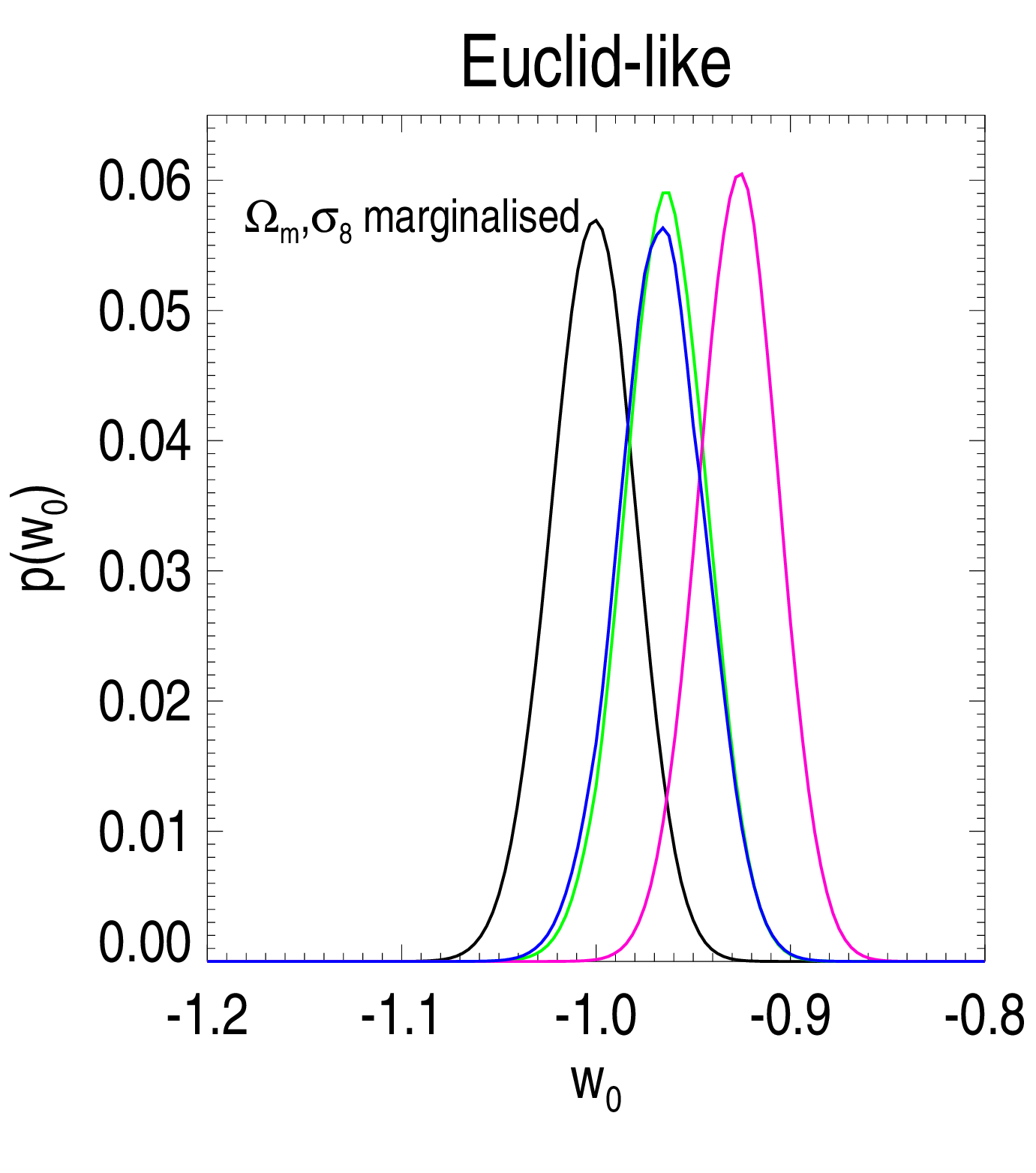,width=0.33\textwidth}
\end{tabular} 
\caption{\label{fig:like_toy}Left panel: joint constraints on
  $\Omega_{\rm m}$ and $\sigma_8$ for a Euclid-like survey, using our
  improved halo model tuned to fit the gas and stellar mass fractions predicted by each scenario to compute the likelihood. The contours indicate
  the $68\%$ confidence regions for the various feedback models. Middle
  panel: joint constraints on $\Omega_{\rm m}$ and $w_0$. Right panel: posterior probability distribution of $w_0$ after marginalisation over $\Omega_{\rm m}$ and $\sigma_8$.   When the improved
  model is used to compute the lensing signal, the biases in the
  recovered cosmological parameters are significantly reduced.}
\end{figure*}
Indeed, we remind the reader, that we currently calibrate
the model using scaling relations covering a large range of masses; in
particular down to low-mass haloes.
Figure \ref{fig:halo_masses} shows the contributions to the power
spectrum from various ranges in halo mass. This demonstrates that
haloes with masses $M_{500}>10^{13}\,h^{-1}\,M_\odot $, i.e.\ massive galaxies,
groups and clusters, contribute most on the scales of interest, but
that the properties of halos with masses as low as $10^{12}\,  h^{-1}\,M_\odot$
need to be understood as well.

Currently, scaling relations can be measured observationally  down to virial masses of
a few $10^{13}\,h^{-1}\, M_{\odot}$, albeit only for biased samples.  For those high-mass haloes the parameters
of our model can be constrained directly by observations of groups and
clusters of galaxies, most notably the gas fraction (e.g.,\ Giodini et
al.\ 2009; Sun et al.\ 2009). Although studies of ensemble averaged
properties can also provide useful constraints (Leauthaud et al. 2010;
Rykoff et al. 2008), for masses down to $M_{\rm vir} \sim 10^{12}M_\odot$
observational constraints are expected to remain limited.
 Furthermore, in our model, we made a rather ad-hoc assumption that
the gas that is removed from the central regions is in the
outskirts of the haloes. Although this seems plausible (see
for example Rasheed et al. 2010) it is not clear how accurately  one can observe
the distribution of the gas in the outskirts of the haloes. Because of those observational limitations, we expect that an accurate description of the baryonic feedback will have to rely in part  on simulations. 

 Finally, our findings suggests that a model with  few parameters, similar to the one we introduced,  can capture the main features of  the baryonic feedback. Those parameters  need to be varied together with the cosmological parameters when one performs a cosmological interpretation.  One can use observations and simulations to derive priors on those parameters and increase the accuracy of the cosmological parameters estimation.

\subsection{Improved constraints on cosmological parameters}

To examine whether the biases on the cosmological parameters are
indeed reduced, we repeat the likelihood analysis described in
Section~\ref{sec:results}, using the improved halo model to compute
the lensing signal. The results are presented in Figure
\ref{fig:like_toy}. For the AGN scenario, where the biases in
$\Omega_{\rm m}$ and $w_0$ were $\sim 20\%$ and $40\%$ respectively, the
bias is now visible reduced, although it is still larger than the
statistical error for a Euclid-like survey.

Despite its simplicity, the model allows for a significant improvement
and its accuracy can be further increased. For instance, we did not
include any evolution in the scaling relations. We also did not
account for the intrinsic dispersion in the model
parameters. Furthermore, for the gas that has been ejected beyond
$r_{500}$, we simply assumed an homogeneous distribution, but more
work is needed to validate (or improve) this assumption. We will
explore various improvements to our model in future work, as they are
beyond the scope of this paper.

\section{Consequences for other parameters}
\label{sec:parameters}

In section \ref{sec:results} we showed that ignoring the effects of
baryons on the power spectrum leads to significant biases in the
recovered dark energy equation of state $w_0$, as well as in $\sigma_8$
and $\Omega_{\rm m}$. The measurement of these parameters is, however,  not
the only goal of future cosmic shear studies. In this section we
examine the impact of baryon physics on the accuracy of some cosmological parameters for which weak lensing has been suggested to be a useful probe.
These are the scale dependence of the spectral index  of the
primordial power spectrum  (section ~\ref{sec:running}), the properties of cold dark matter  (section~\ref{sec:wdm})  and the neutrino masses (section~\ref{sec:neutrino}).

\subsection{Running of the spectral index}\label{sec:running}

The  large-scale structure we observe today is believed
to arise from quantum fluctuations that grew to cosmological scales
during a period of inflation. Simple models of inflation predict a
primordial power spectrum that is nearly scale invariant, $P(k)\propto
k^{n_s}$, with scalar spectral index $n_s\approx 1$. In more
complicated inflation theories $n_s$ may also depend on $k$. A general
expansion adopted for $n_s(k)$ is \cite{KoTu95}:

\be
n_s(k)=n_s(k_0)+\frac{dn_s}{d\ln k}\ln\Big(\frac{k}{k_0}\Big) 
\ee 

\noindent where $k_0=0.015 ~{\rm Mpc}^{-1}$ is a pivot. A non-zero value
for $\alpha_s\equiv{d n_s}/{d \ln k}$ results in a running spectral
index. Measurements of both $n_s$ and $\alpha_s$ provide direct
constraints on the inflationary models, and  are thus of great
interest. CMB temperature fluctuations as well as CMB polarisation
measurements are able to provide constraints on the spectral index. The
most recent results from  ACT  WMAP7 suggest $n_s(k_0)$ to be
slightly smaller than  unity and $\alpha_s<0$ \cite{Duetal10}.

\begin{figure}
\psfig{figure=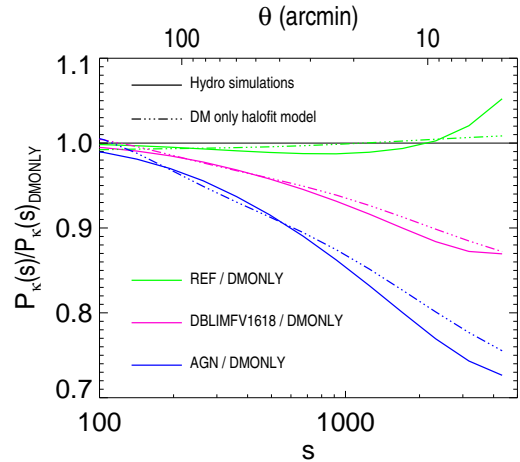,width=.45\textwidth}
\caption{\label{fig:2pt_tomo_running} Ratio of the weak lensing power
  spectra for the lowest redshift tomographic bin, for the
  REF/DMONLY, DBLIMFV1618/DMONLY and AGN/DMONLY models (solid lines).
  The dot-dashed lines show the ratio for the dark matter only halofit  model that
  best fits each of the simulations, obtained by varying $\sigma_8$,
  the spectral index $n_s(k_0)$ and the running $\alpha_s={d n_s}/{d
    \ln k}$. In general it is possible to find a combination of
  parameters, such that the dark matter only model with a running of the spectral
  index fits the simulated power spectra reasonably well.}
\end{figure}

Weak lensing observations enable us to extend the measurements to
smaller scales, thus tightening the constraints on $\alpha_s$.  We are
concerned, however, that such a measurement will also be  affected by baryon
physics, because the suppression in the power spectrum seen for the
AGN model could also be achieved by a more negative value for
$\alpha_s$.  This is demonstrated by Figure
\ref{fig:2pt_tomo_running}, where we show that it is possible to find
dark matter only models (i.e. halofit models) with a running spectral index that match the
simulated power spectra for the lowest redshift bin. 

\begin{figure*}
\begin{center}
\psfig{figure=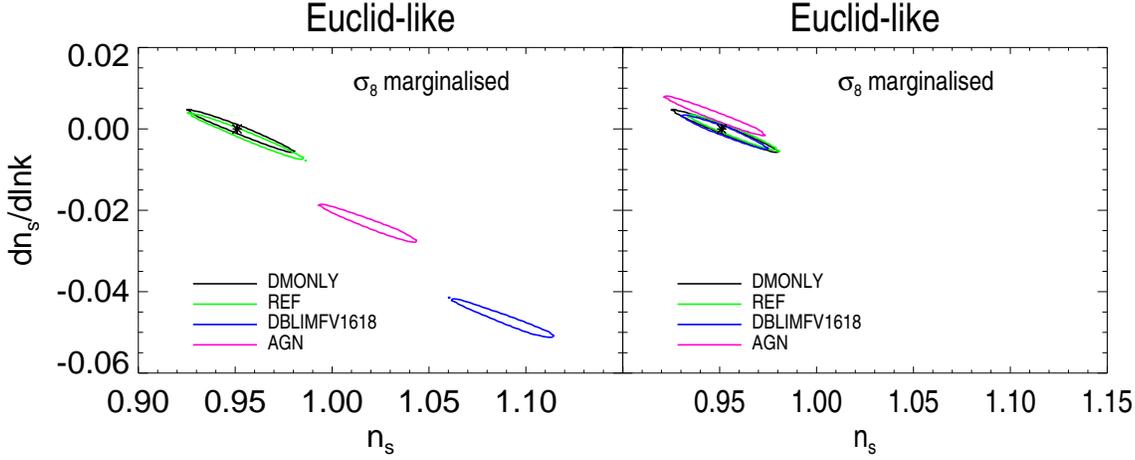,width=.90\textwidth}
\end{center}
\caption{\label{fig:running} Left panel: joint constraints on the
  primordial spectral index $n_s(k_0)$ and running $\alpha_s={d
    n_s}/{d \ln k}$ after marginalising over $\sigma_8$ (while keeping
  the other parameters fixed) for a Euclid-like survey. The contours
  indicate the 68\% confidence regions for the feedback models that
  are indicated. Right panel: constraints when we use our extension of
  the halo model (see Section~\ref{sec:toy_model}), which was tuned to fit the halo gas fractions and $K$-band magnitudes predicted by the simulations. The improved  halo model partially reduces the
  level of the bias.}
\end{figure*}

We also computed
the joint likelihood for $n_s \in [0.90,1.20] $ and $\alpha_s \in [-0.10, 0.050]$, marginalising over a flat prior 
$\sigma_8 \in [0.65,0.83]$ while keeping $w_0$ and $\Omega_{\rm m}$ fixed.     The results are
presented in the left panel of Figure~\ref{fig:running}. As expected,
the best-fit value for $\alpha_s$ is biased to a more negative value
in the case of the AGN model if the effects of baryons are
ignored. Our simple extension of the halo model is able to
reduce  the bias also  for these parameters, as is
demonstrated by the right panel in Figure~\ref{fig:running}.

\subsection{Warm dark matter}\label{sec:wdm}

A number of observations appear to disagree with the outcome of
 $\Lambda{\rm CDM}$ simulations which predict a large amount of structure in galaxy-sized haloes.  In particular, the
simulations predict a larger number of satellite galaxies for the
Milky Way than is observed (the so-called missing satellites problem),
although the recent discovery of faint satellites appears to alleviate this problem (see Belokurov et al.\ 2010). Another longstanding
issue concerns the inner slope of the density profile: low surface brightness galaxies show profiles which are not as cuspy
as  the NFW profile (e.g. Kuzio De Naray et al.\ 2010; Oh et al.\ 2010). 

The existence of warm dark matter (WDM)  can, in principle, solve the problems
mentioned above, because the free-streaming of the particles
suppresses small-scale fluctuations and leads to haloes with a core. We
note, however, that Villaescusa \& Dalal (2011) recently showed that 
WDM  haloes cannot form cores that are large enough to match
the observations. Recent papers suggest that future weak lensing missions such as Euclid, combined
with Planck observations, can constrain the mass of the WDM particles,
through the effect they have on the growth of structure (e.g., Markovi\v{c} et al.\ 2010).

To ensure that structure formation is only altered on small scales,
the parameters of WDM models are generally chosen such that that the
power spectrum is modified at subgalactic scales $k_{\rm fs} \simeq 0.1-1
h~{\rm Mpc}^{-1}$, which are also affected by baryon physics. As an
example, we consider the case of fully thermalised particles described
by Markovi\v{c} et al.\ (2010).  In this case the WDM streams freely
shortly before the Universe enters the matter-dominated era. The
effect on the power spectrum can be parametrised by (Viel et al.\ 2005):

\be
T(k)=\Big(\frac{P_{\rm WDM}(k)}{P_{\rm CDM}(k)}\Big)^{1/2} = [
  1+(\alpha k)^{(2 \mu)}]^{(-5/\mu)} ,
\ee 

\noindent with $\mu=1.12$ and where $P_{\rm CDM}(k)$ and $ P_{\rm
  WDM}(k)$ are the CDM and WDM power spectra, respectively.  The
value for the break scale $\alpha$ is then given by (Viel et al.\ 2005):

\be
\alpha= 0.049 \Big(\frac{m_{\rm WDM}}{\rm keV}\Big)^{-1.11}
\Big(\frac{\Omega_{\rm WDM}}{0.25}\Big)^{0.11}
\Big(\frac{h}{0.7}\Big)^{1.22} h ^{-1} {\rm Mpc}, 
\ee

\noindent where $\Omega_{\rm WDM}$ is the WDM density parameter and
$m_{\rm WDM}$ is the WDM particles mass. Observations of the power
spectrum of the Lyman-$\alpha$ forest suggest $m_{\rm WDM}> 4{\rm
  keV}$ \cite{Seetal06,Vietal08,Boetal09}. If we use this limit, we find that
for $k=10\,h\,{\rm Mpc}^{-1}$, which is where the suppression of the
total power by baryons is maximal (see Fig.~\ref{fig:Pkappa}), the
power spectrum is suppressed by only $5\%$, which is much smaller than
the suppression predicted by the AGN feedback model. The suppression
decreases rapidly, being less than a percent for $k=1\, h~{\rm
  Mpc}^{-1}$. One could argue that this result, which has been derived
in the linear approximation, is not valid at small scales where the
power spectrum is in the non-linear regime. However, Boehm et
al.\ (2005), showed that the non-linear evolution reduces the
difference between CDM and WDM models (also see Markovi\v{c} et
al.\ 2011).

We therefore conclude that the effect of WDM on the weak lensing
signal is likely to be dwarfed by baryon physics. If we wish to
improve upon current constraints on the WDM mass, we would need to be
able to compute the effects of baryon physics on the total matter
power spectrum to less than a percent for $k \gsim 1\, h\,{\rm Mpc}^{-1}$.
It is not clear that such accuracy can be achieved.

\subsection{Massive neutrinos}\label{sec:neutrino}

Particle physics experiments have conclusively shown that neutrinos
have mass and suggest that there are three species. Experiments are also
able to measure the differences in the square of the masses, and thus
provide a lower bound on the mass of the heaviest neutrino species.
The effect of massive neutrinos on the power spectrum is similar to
that of warm dark matter. The main difference is that
neutrinos  are much lighter (i.e, they are considered hot
dark matter). As a consequence, they become non-relativistic in the
matter dominated era.

While experiments are designed to measure the mass of the electron
neutrino, studies of large-scale structure  constrain the total
density of neutrinos in the universe. The most recent estimate from
WMAP7 is  $\sum m_\nu<1.3\, {\rm eV}$, or $\sum
m_{\nu} <0.65\, {\rm eV}$ if constraints from baryon acoustic
oscillations are included \cite{Koetal11}. This yields an upper bound
for on the total neutrino density of $\Omega_{\nu}<0.02$ for WMAP7, where
we used the relation between the neutrino mass and their cosmological
density:

\be \Omega_\nu= \frac{\sum m_\nu}{93.8 h^2 {\rm eV}}\,. \ee 

As before, massive neutrinos do not modify the power spectrum on
scales larger than their free-streaming length, which decreases with
time. Therefore there is a $k_{\rm nr}$ below which the power
spectrum is not affected \cite{LePa06}: 

\be 
k_{\rm nr}=0.018\,\Omega_{\rm m}^{1/2}\Big(\frac{m_\nu}{1 {\rm eV}} \Big)^{1/2}\,. 
\ee

\begin{figure}
\psfig{figure=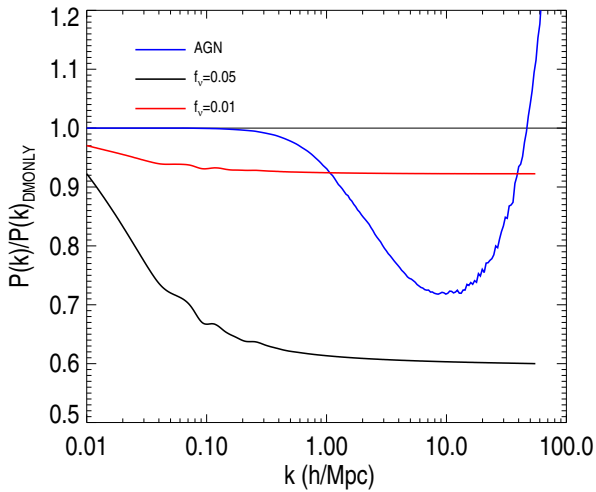,width=.45\textwidth}
\caption{\label{fig:P_neutrino} Ratio of the AGN/DMONLY power spectra
  (blue line), and dark matter power spectra with
  $f_\nu\equiv\Omega_\nu/\Omega_{\rm m}=0.01$ and $0.05$, which correspond to
  neutrino masses of $\sum m_\nu\sim 6.0$ and $\sum m_\nu\sim 1.2$ eV, respectively. The effect of massive neutrinos on the power
  spectrum is quite different from that of baryon physics, even if
  neutrinos are light.}
\end{figure}

\noindent To compute the effect of neutrinos on the power spectrum, we
use the transfer functions given by the Boltzmann code CAMB (Lewis et al.\ 2000). Figure \ref{fig:P_neutrino} shows the ratio of power
spectra at $z=0.5$ with neutrino fractions $f_\nu \equiv
\Omega_\nu/\Omega_{\rm m} =0.01$ and
$f_\nu=0.05$. Note that the best-fit WMAP7 cosmology \cite{Koetal11}  yields $k_{\rm nr} \lesssim 0.01~ h~{\rm Mpc}^{-1}$ and  $f_\nu <0.1$.

At intermediate scales our results indicate that neutrinos cause a
scale-dependent suppression of the power spectrum. For large $k$, our
calculations, which are based on linear perturbation theory, indicate
that the ratio $P_\nu(k)/P_{\rm CDM} (k)$ is (almost) constant with a
value $\simeq (1-8 f_{\nu})$ \cite{EiHu98}. A concern might be that linear theory does not
provide an adequate description. Although the highly non-linear regime
has not been studied, the numerical simulations presented in Brandbyge et al.\ (2008) suggest that our results are sufficiently accurate:
$P_\nu(k)/P_{\rm CDM}(k)$ reaches a minimum value of $\sim (1- 10
f_{\nu})$ and slightly increases again for smaller $k$. The results
presented in Figure~\ref{fig:P_neutrino} indicate that for $f_\nu \sim
0.05$ (or a mass of $ \sum m_\nu\sim 6.0$~eV) the suppression of power by massive
neutrinos dominates over the baryon physics. In the case of light
neutrinos ($\sum m_\nu\sim 1.2$~eV or $f_\nu=0.01$), baryon physics dominates
for $k>1\, h\, {\rm Mpc}^{-1}$, but the neutrinos cause a significant
suppression of power on intermediate scales.  Thus, it seems likely that one would be able to get useful constraints using only large-scale measurements.

\section{Conclusions}\label{sec:conclusions}

Weak gravitational lensing by large-scale structure is a powerful
technique to study the properties of dark matter and dark energy.
However, the interpretation of the signal requires comparison to a
predicted matter power spectrum. To date, such cosmic shear
measurements have used predictions based on numerical simulations of
cold dark matter. As the precision of future projects improves, the
effects of baryon physics need to be quantified.  Recently, van Daalen
et al. (2011) have shown that the impact of baryonic feedback is
significant and demonstrated that the precision with which one is able
to predict the matter power spectrum from simulations depends on how
well one understands and models mechanisms such as radiative cooling,
star formation, metal enrichment, supernovae and AGN feedback. In
particular, they have shown that baryon physics modifies the matter
power spectrum to a level which is higher than the accuracy required
for future weak lensing experiments, which require predictions for the
amplitude of the power spectrum with an accuracy of a few percent for a
large range of scales.

In this paper we made a first attempt to quantify the effect of
baryon physics on the cosmic shear signal, using a subset of the
cosmological hydrodynamic simulations that were carried out as part of
the OWLS project (Schaye et al.\ 2010) and for which the matter power
spectra have been measured by van Daalen et al (2011).

We find that if weak lensing measurements are interpreted using a
power spectrum that ignores the effects of baryon physics, the
resulting cosmological parameter estimates are biased. The bias
depends on the feedback model, but is as large as $\sim 40\%$ for the
dark energy equation of state $w_0$ for the AGN scenario, which we
consider our most realistic model as it reproduced the observed X-ray
and optical properties of groups of galaxies (McCarthy et
al. 2010). The biases dominate over the statistical errors for Euclid,
a proposed space mission that aims to measure $w_0$ with a precision
of a few percent.

The reason for the large bias is the fact that efficient feedback
modifies the power spectrum out to several megaparsecs, contrary to
the commonly held belief that only very small scales are affected, as
is the case in previous models that suffered from the ``overcooling''
problem. To reduce the bias to acceptable levels, one would need to
discard scales smaller than $\sim 40$ arcminutes from the weak lensing
analysis in the case of Euclid, at the expense of a factor three
increase in the statistical error.

Such a large reduction in sensitivity to the dark energy properties is
not desirable. Instead, we explored an approach to reduce the bias,
based on a simple extension of the halo model. We treat the dark
matter, gas and stars as separate components, assigning 
suitable profiles to each.  The baryonic mass assigned to each component is
determined using scaling relations between the halo mass and the gas
fraction and stellar mass. For each scenario the model
parameters were constrained by fitting the simulated gas fraction and
stellar masses of groups and clusters at redshift $z=0$. We also made
the ad-hoc assumption that the ejected gas is distributed uniformly
between the virial radius and $2r_{\rm vir}$. Further work is needed
to improve upon this assumption.

Although several improvements should be made (an issue that we will
explore in future work), the simple model presented here is able to
describe the OWLS power spectra fairly well for $k< 10\, h\,{\rm
  Mpc}^{-1}$. Particularly encouraging is the reduction in the bias in
$w_0$ and the other cosmological parameters. 

The usefulness of our model hinges partly on our ability to calibrate
the model though observations of the distribution of gas and stars in
halos with masses as low as $10^{12}\,  h^{-1}\, M_\odot$, which may be too
low for X-ray and SZ observations. Moreover, in order to achieve the
precision required for future weak lensing experiments, one would need
to know the distribution of the gas in the peripheral regions of
haloes.  This suggests that improvements require a `hybrid' approach,
relying on a combination of deeper observations and better
simulations. For instance, dedicated X-ray and SZ studies can help
calibrate the model parameters, but they also provide constraints on
the accuracy of the hydrodynamic simulations.

Finally, we also examined the effects of baryon physics on other
applications of weak lensing.  For Euclid, constraints on the
parameters of the primordial power spectrum predicted by theories of
inflation are also significantly biased. The measurement of the
neutrino mass is also affected (in particular if the mass is small),
but the bias should be small if only scales larger than $\sim 10$ Mpc
are considered.  Improved weak lensing constraints on the properties
of warm dark matter are unlikely, because current constraints lead to
changes in the power spectrum that are much smaller than those
introduced by baryon physics.

\section*{Acknowledgments}

The authors wish to thank Craig Booth, Alexis Finoguenov, Stefania Giodini, Ben Oppenheimer,  Aaryn Tonita, Edo van Uitert and Malin Velander for useful discussions.  ES, HH and JS acknowledge
support from the Netherlands Organization for Scientific Research
(NWO) through VIDI grants. HH also acknowledge support from a Marie Curie International
Reintegration Grant. This work has been supported by the Marie Curie Training Network CosmoComp (PITN-GA-2009-238356).

\end{document}